\documentclass[twocolumn]{aastex631}
 
 \usepackage{amssymb}
\usepackage{amsmath}
\usepackage{hyperref}
\usepackage{natbib}
\usepackage{color}
\usepackage{graphicx}
\usepackage{amsmath}
\usepackage{bm}

\usepackage{amsmath}

\shorttitle{Transition from Imbalanced to Balanced Kinetic Alfv\'enic Turbulence}
\shortauthors{Zhao et al.}

\graphicspath{{./}{figures/}}

\begin{document}


\title{Observations of Transition from Imbalanced to Balanced Kinetic Alfv\'enic Turbulence}

\author{Jinsong Zhao}
\affiliation{Purple Mountain Observatory, Chinese Academy of Sciences, Nanjing 210023,  People's Republic of China}

\author{Trevor A. Bowen}
\affiliation{Space Sciences Laboratory, University of California, Berkeley, CA 94720-7450, USA}

\author{Stuart D. Bale}
\affiliation{Physics Department, University of California, Berkeley, CA 94720-7300, USA}
\affiliation{Space Sciences Laboratory, University of California, Berkeley, CA 94720-7450, USA}

\author{Chen Shi}
\affiliation{School of Physics and Electronic Sciences, Changsha University of Science and Technology, Changsha, China}

\author{Thierry Dudok de Wit}
\affiliation{LPC2E, CNRS, CNES, University of Orl\'eans, Orl\'eans, France}
\affiliation{International Space Science Institute (ISSI), Bern, Switzerland}

\author{Nikos Sioulas}
\affiliation{Space Sciences Laboratory, University of California, Berkeley, CA 94720-7450, USA}

\correspondingauthor{Jinsong Zhao}
\email{zhaojs82@gmail.com}


\begin{abstract}

We report observations of solar wind turbulence derived from measurements by the Parker Solar Probe. Our findings reveal the emergence of finite magnetic helicity within the transition range of the turbulence, aligning with signatures of kinetic Alfv\'en waves (KAWs). Notably, as the wave scale transitions from super-ion to sub-ion scales, the ratio of KAWs with opposing signs of magnetic helicity initially increases from approximately 1 to 6.5 before returning to 1. This observation provides, for the first time, compelling evidence for the transition from imbalanced kinetic Alfv\'enic turbulence to balanced kinetic Alfv\'enic turbulence.

\end{abstract}


\section{Introduction} \label{sec:intro}

Alfv\'enic turbulence is prevalent in the heliosphere \citep{tu1995mhd,bruno2013solar} and is believed to exist in astrophysical environments \citep{armstrong1995electron,goldreich1995toward}. The dissipation of this turbulence is proposed to be a major contributor to heating in the solar wind, solar corona, and other astrophysical plasmas \citep{quataert1999turbulence,2006ApJ...645L..85S,cranmer2007self,van2011heating,chandran2010perpendicular,chandran2010alfven,kawazura2019thermal,arzamasskiy2019hybrid,bowen2020constraining,Wu2020ApJ,Wu2021ApJ}. Observations indicate that Alfv\'enic turbulence encompasses a broad range of scales, evolving from magnetohydrodynamics (MHD) scales down to sub-ion kinetic scales, potentially extending to sub-electron kinetic scales  \citep{leamon1998observational,bale2005measurement,alexandrova2009universality,sahraoui2009evidence,sahraoui2010three,salem2012identification,chen2013nature}. As the turbulence cascade scale approaches the ion gyroradius, anisotropic turbulence models predict a transition from MHD-scale Alfv\'enic waves to kinetic Alfv\'en waves (KAWs) \citep{quataert1998particle,howes2008model,schekochihin2009astrophysical,voitenko2011turbulent,boldyrev2012spectrum,zhao2013kinetic,zhao2016kinetic,chen2017nature,Schekochihin2022JPlPh}. Despite substantial evidence of KAWs occurring at ion kinetic scales in satellite observations and numerical simulations \citep{bale2005measurement,alexandrova2009universality,sahraoui2009evidence,sahraoui2010three,salem2012identification,chen2013nature,howes2011gyrokinetic,2015ApJ...812...21F,grovselj2018fully,grovselj2019kinetic,squire2022high}, the precise evolution of turbulence at ion kinetic scales remains unknown.

Solar wind turbulence is often highly imbalanced at MHD scales in the inner heliosphere, with outward-propagating fluctuations away from the Sun prevailing over inward-propagating ones \citep{tu1995mhd,Chandran2008ApJ,chen2013residual,yang2023energy}. Based on the signatures of magnetic helicity at ion scales observed in solar wind turbulence \citep{he2011possible,podesta2011magnetic,Telloni2015ApJ,Telloni2020ApJ}, it has been suggested that KAWs are predominantly outward-propagating, indicating that imbalanced kinetic Alfv\'enic turbulence arises at ion kinetic scales \citep{podesta2011magnetic,He2012ApJ,Klein2014ApJ}. The evolution of such imbalanced turbulence may be influenced by nonlinear interactions among co-propagating KAWs or by mechanisms involving the helicity barrier \citep{voitenko2016mhd,meyrand2021violation,squire2022high,McIntyre2025PhRvX,Panchal2025ApJ}. Moreover, at kinetic scales, the power spectral density of the magnetic field is often characterized by a transition range, occurring at scales comparable to the ion scale, and a kinetic-inertial range between ion and electron scales \citep{bale2005measurement,alexandrova2009universality,sahraoui2009evidence,sahraoui2010three,salem2012identification,chen2013nature,2020ApJS..246...55D,duan2021anisotropy,huang2021ion}. These observations prompt a fundamental question concerning the evolution of turbulence at kinetic scales: how does imbalanced Alfv\'enic turbulence evolve toward smaller scales? This phenomenon remains largely unexplored experimentally, primarily due to the absence of reliable diagnostic techniques capable of quantifying wave propagation direction at ion and sub-ion kinetic scales in single-point spacecraft observations.

Using in-situ measurements from Parker Solar Probe \citep[PSP;][]{bale2016fields,kasper2016solar}, this Letter investigates solar wind turbulence at kinetic scales through a novel magnetic helicity-based method capable of quantifying wave propagation imbalance at kinetic scales. Our findings present, for the first time, observational evidence of the transition from imbalanced to increasingly balanced kinetic Alfv\'enic turbulence in the solar wind, offering new insights into the evolution of turbulence imbalance. The observed imbalance development agrees with key predictions of the helicity barrier model \citep{meyrand2021violation,squire2022high}, linking our measurements directly to a proposed cascade mechanism near ion kinetic scales.

\section{Methodology and data} 
\label{sec:overview}

Quantifying the imbalance in kinetic Alfv\'enic turbulence at ion and sub-ion scales requires determining the propagation direction of turbulent fluctuations relative to the background magnetic field ${\boldsymbol B}_0$. This propagation information can be extracted from the phase coherence between parallel and perpendicular magnetic field perturbations (see Appendix A), characterized by the magnetic helicity $\sigma_\mathrm{m\perp_2\parallel}$ (see Equation (2) for definition). An accurate diagnosis requires a field-aligned coordinate system, with axes defined by $\hat{\boldsymbol e}_{\perp_1}={\boldsymbol k}_\perp/k_\perp$, $\hat{\boldsymbol e}_{\perp_2} = \hat{\boldsymbol e}_\parallel \times \hat{\boldsymbol e}_{\perp_1}$, and ${\boldsymbol e}_\parallel \equiv {\boldsymbol B}_0/|{\boldsymbol B}_0|$ constructed from ${\boldsymbol B}_0$ and the wave vector ${\boldsymbol k} = {\boldsymbol k}_\perp + {\boldsymbol k}_\parallel$, where ${\boldsymbol k}_\perp$ and ${\boldsymbol k}_\parallel$ are the components perpendicular and parallel to ${\boldsymbol B}_0$, respectively. In this Letter, we develop a diagnostic method within this coordinate system to accurately resolve wave propagation properties in kinetic-scale turbulence. Details of the implementation and the improvements over previous methods are provided in Appendix A.

A central feature of our method is determining the ${\boldsymbol k}_\perp$ direction by exploiting a characteristic property of KAWs: their magnetic perturbations exhibit highly elliptical or linear polarization within the plane perpendicular to ${\boldsymbol B}_0$ \citep{Hollweg1999JGR,Zhao2014ApJ,Zhao2022ApJ}. The dominant component of the magnetic perturbation lies perpendicular to both ${\boldsymbol B}_0$ and ${\boldsymbol k}_\perp$, allowing ${\boldsymbol k}_\perp$ to be inferred by locating the dominant magnetic perturbation (see Appendix A for implementation details). Once ${\boldsymbol k}_\perp$ and the field-aligned coordinate system are determined, the wave propagation direction is inferred from the sign of $\sigma_{\mathrm{m\perp_2\parallel}}$, where positive values indicate propagation along ${\boldsymbol B}_0$ and negative values indicate propagation counter to ${\boldsymbol B}_0$. Notably, this signature remains consistent in both the solar wind and spacecraft frames.

Because sub-ion Alfv\'enic turbulence in the near-Sun solar wind extends to frequencies exceeding 100 Hz \citep{bowen2020constraining}, we analyze magnetic turbulence imbalance from 09:30 UTC on 2018 November 04 to 21:20 UTC on 2018 November 07, using high-cadence ($\sim$ 293 Hz) magnetic field data merged from fluxgate and search coil magnetometers on PSP/FIELDS \citep{bowen2020merged}. Proton plasma parameters are obtained from Solar Probe Cup (SPC) measurements \citep{case2020solar} on PSP/SWEAP \citep{Kasper2016SSRv}, and electron parameters are derived from fitting the quasi-thermal noise measurements \citep{Moncuquet2020ApJS}. Because the cadence of PSP/SPC is lower than $\sim$1 Hz, the imbalance of sub-ion Alfv\'enic turbulence in the interval of interest cannot be tracked using the conventional cross-helicity parameter.

We apply the Morlet wavelet transform to the time series magnetic field data to obtain the complex wavelet amplitude ${\boldsymbol W}(f,t)$, where the central frequency at each scale $s$ is given by $f=(w_0+\sqrt{2+w_0^2})/(4\pi s)$, with a nondimensional frequency parameter $w_0 = 6$ \citep{Torrence1998BAMS}.
From these, we calculate the normalized reduced magnetic helicity defined as follows:
\begin{eqnarray}
\sigma_{\mathrm{m\perp_1\perp_2}}    &=& -2 \mathrm{Im}(W_{\perp_1}W^{*}_{\perp_2})/|{\boldsymbol W}|^2, \\
\sigma_{\mathrm{m\perp_2\parallel}}  &=& -2 \mathrm{Im} (W_{\perp_2}W^{*}_{\parallel})/|{\boldsymbol W}|^2, \\
\sigma_{\mathrm{m\parallel\perp_1}} &=& -2 \mathrm{Im}(W_{\parallel} W^{*}_{\perp_1})/|{\boldsymbol W}|^2,
\end{eqnarray}
where ``$*$'' denotes the complex conjugate and $|{\boldsymbol W}|^2= |W_{\perp_1}|^2 + |W_{\perp_2}|^2 + |W_{\parallel}|^2$.

Among these helicity components, $\sigma_{\mathrm{m\perp_2\parallel}}$ quantifies the phase coherence between $W_{\perp_2}$ and $W_{\parallel}$, serving as a clear signature of KAWs. By comparison, $\sigma_{\mathrm{m\perp_1\perp_2}}$ captures the coherence between $W_{\perp_1}$ and $W_{\perp_2}$, and is commonly used for diagnosing quasi-parallel ion cyclotron waves (ICWs) and fast-magnetosonic whistler waves (FMWs) \citep{bowen2020ion, verniero2020parker, liu2023radial, 2024ApJ...971L..41S}. However, both quasi-perpendicular KAWs and quasi-parallel ICWs/FMWs generally exhibit weak coherence between $W_{\perp_1}$ and $W_{\parallel}$, so this helicity component is not a reliable diagnostic for any of these wave modes.


\section{Event overview} 

\begin{figure}[t]
\centering
\includegraphics[width=3.5in]{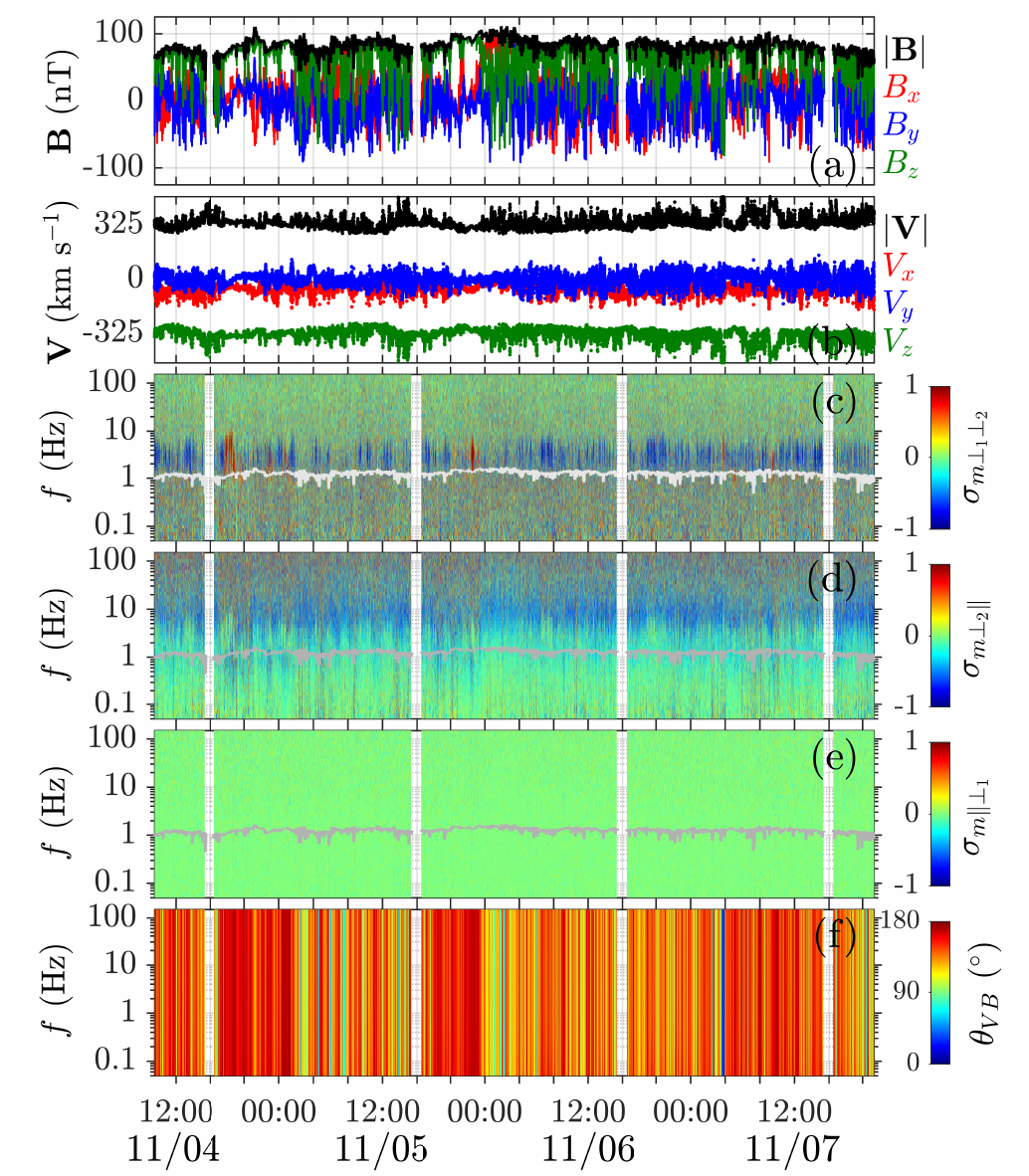}
\caption{ Plasma environment and magnetic helicity during the time interval from 09:30 UTC on 2018 November 04 to 21:20 UTC on 2018 November 07. (a) The magnitude and three components of the magnetic field ${\boldsymbol B}$ in the spacecraft frame.  (b) The solar wind speed ${\boldsymbol V}$ in the spacecraft frame. (c) The distribution of $\sigma_\mathrm{m\perp_1\perp_2}$. (d) The distribution of $\sigma_\mathrm{m\perp_2 \parallel}$. 
(e) The distribution of $\sigma_\mathrm{m \parallel \perp_1}$. (f) The distribution of $\theta_\mathrm{VB}$. The gray curves in (c)--(e) denote the proton cyclotron frequency $f_{\mathrm{cp}}$. Because the merged magnetic field data in (a) contain gaps, no corresponding magnetic helicity or $\theta_{\mathrm{VB}}$ distributions are available in (c)–(f). }
    \label{fig:1}
\end{figure}

Figure~\ref{fig:1} provides an overview of the plasma environment and the normalized magnetic helicity distributions during the chosen time interval.

Figures \ref{fig:1}(a) and (b) display the measured magnetic field and solar wind velocity in the spacecraft frame, with $\hat{\boldsymbol e}_z$ oriented sunward. The magnetic field strength remains relatively constant, while its direction exhibits  considerable variability due to the presence of switchbacks. In contrast, the solar wind velocity experiences only minor changes, predominantly streaming outward from the Sun along the radial direction.

Figure \ref{fig:1}(c) shows the distribution of $\sigma_\mathrm{m\perp_1\perp_2}$, which quantifies the polarization of the waves relative to the magnetic field. Positive (red) and negative (blue) $\sigma_\mathrm{m\perp_1\perp_2}$ correspond to right- and left-handed polarized waves, respectively. The data show a predominance of left-handed polarized waves with $\sigma_\mathrm{m\perp_1\perp_2} \rightarrow -1$ at $f \gtrsim f_{\mathrm{cp}}$, where $f_{\mathrm{cp}}$ is the proton cyclotron frequency. Intermittent right-handed polarized waves with $\sigma_\mathrm{m\perp_1\perp_2} \rightarrow 1$ are also present. These polarization characteristics are consistent with the presence of ICWs and FMWs  \citep{bowen2020ion, verniero2020parker, liu2023radial}.

Figure~\ref{fig:1}(d) presents the distribution of $\sigma_{\mathrm{m\perp_2\parallel}}$, which is used to identify KAWs, as described in the Appendix C. A clear signature of $\sigma_{\mathrm{m\perp_2\parallel}} < 0$ appears at frequencies $f \gtrsim 1$ Hz, indicating the presence of KAWs in the near-Sun solar wind, consistent with previous observations \citep{huang2020kinetic,duan2021anisotropy}. Furthermore, the signature of negative $\sigma_{\mathrm{m\perp_2\parallel}}$ varies over time. This temporal variation is closely linked to the angle $\theta_{\mathrm{VB}}$ between the local solar wind magnetic field and velocity direction (Figure~\ref{fig:1}(f)), as shown by the joint distributions of $\sigma_{\mathrm{m\perp_2\parallel}}$ and $\theta_{\mathrm{VB}}$ in Figure~\ref{fig:2}(a).

Figure \ref{fig:1}(e) illustrates the distribution of $\sigma_\mathrm{m \parallel \perp_1}$. This quantity is typically distributed around zero, suggesting a lack of coherence between $W_{\perp_1}$ and $W_{\parallel}$. This behavior aligns with theoretical predictions that $\sigma_\mathrm{m\parallel \perp_1} \sim 0$ for linear KAWs and quasi-parallel coherent ICWs and FMWs.

As revealed by the helicity distributions in Figures \ref{fig:1}(c)–(d), both coherent wave signatures and broadband turbulent fluctuations appear in the observations. To accurately investigate the properties of the turbulence itself, it is essential to isolate intervals dominated by turbulent fluctuations. To achieve this, we exclude time intervals containing coherent waves, identified by significant $\sigma_\mathrm{m\perp_1\perp_2}$ values, based on two criteria: (1) the ratio of the upper to lower boundary frequencies of the relevant frequency band exceeds 2; and (2) the maximum value of $|\sigma_\mathrm{m\perp_1\perp_2}|$ exceeds $3\sigma_\mathrm{std}$, where $\sigma_\mathrm{std}$ denotes the mean standard deviation of $\sigma_\mathrm{m\perp_1\perp_2}$ within the low-frequency range of 0.08--0.8~Hz. Additionally, we remove adjacent time intervals that are potentially affected by these coherent waves. The resulting occurrence rate of the identified turbulent fluctuations is approximately 40\%. Details of the identification procedure are provided in Appendix B.

\section{ Magnetic helicity and turbulent spectra } 
 
\begin{figure}[t]
    \centering
    \includegraphics[width=3.5in]{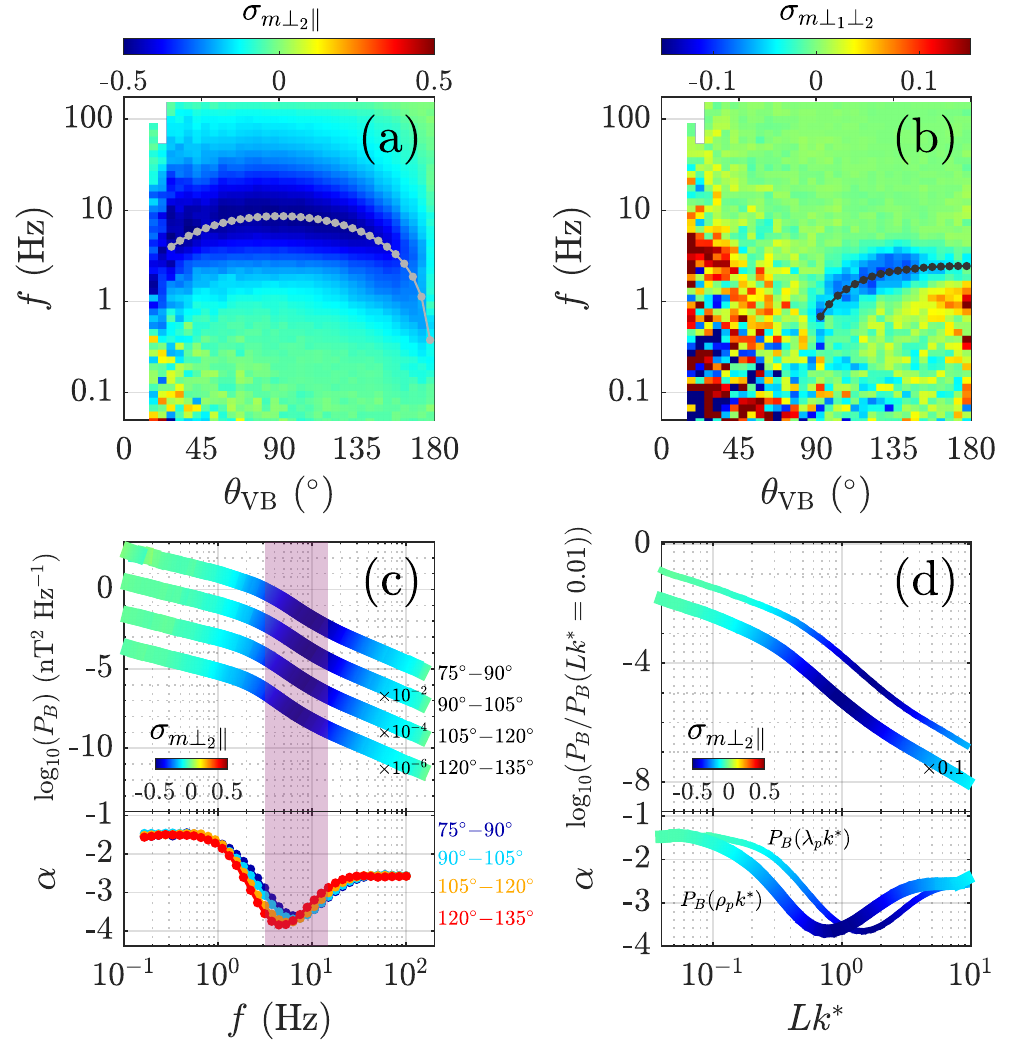}
   \caption{(a) The $\sigma_{\mathrm{m\perp_2\parallel}}$ distribution in the $\theta_{\mathrm{VB}}$--$f$ space, where the gray curve denotes the Doppler shift frequency of quasi-perpendicular waves with $\rho_pk=1$. 
   (b) The $\sigma_{\mathrm{m\perp_1\perp_2}}$ distribution in the $\theta_{\mathrm{VB}}$--$f$ space, where the black curve denotes the Doppler shift frequency of quasi-parallel waves with $\lambda_pk=0.6$ and $\theta=170^\circ$.
 (c) The power spectral density of the magnetic field $P_B$ in four different $\theta_{\mathrm{VB}}$ ranges: $75^\circ$--$90^\circ$, $90^\circ$--$105^\circ$, $105^\circ$--$120^\circ$, and $120^\circ$--$135^\circ$. 
 (d) The $P_B$ as a function of the normalized wavenumber $Lk^*$, with $L=\rho_p$ (thick curve) and $L=\lambda_p$ (thin curve).   
In (c) and (d), the upper panels show $P_B$ with $\sigma_{\mathrm{m\perp_2\parallel}}$ overlaid, whereas the lower panels display the corresponding spectral indices $\alpha$. 
  }
    \label{fig:2}
\end{figure}

A crucial parameter influencing the distributions of reduced magnetic helicity and magnetic power spectral density ($P_B$) is the angle $\theta_{\mathrm{VB}}$ \citep{he2011possible,podesta2011magnetic}. The relationships among these parameters for identified turbulent fluctuations are illustrated in Figure \ref{fig:2}. To choose a quiet background for analysis, data with an angular deviation greater than $15^\circ$ between ${\boldsymbol B}$ and ${\boldsymbol B_0}$, referred to as $\theta_{\mathrm{BB_0}}$, are excluded from consideration \citep{woodham2021dependence}. Moreover, some plasma measurements have low sampling rates or are sampled over 20 second intervals, and thus, we discard any data with average plasma sampling rates below 1 Hz (i.e., fewer than 20 samples in a 20 second interval).

Figure \ref{fig:2}(a) presents the distribution of $\sigma_{\mathrm{m\perp_2\parallel}}$ as a function of $f$ and $\theta_{\mathrm{VB}}$. A region of strongly negative helicity (blue) is observed, with its characteristic frequency decreasing as $\theta_{\mathrm{VB}}$ deviates from $90^\circ$. 
This frequency dependence provides direct information on the wave normal angle. The observed frequency $f = f_{\mathrm{pl}} + f_{D}$ consists of the wave  frequency $f_{\mathrm{pl}}$ in the solar wind rest frame and the Doppler-shifted  frequency $f_{D} = {\boldsymbol V} \cdot {\boldsymbol k}$.  $f_{D}$ can be further separated into contributions from the parallel and  perpendicular wave vectors: 
$f_\mathrm{Dk_\parallel} = {\boldsymbol V}_\parallel \cdot 
{\boldsymbol k}_\parallel /(2\pi) = (V k_\parallel / 2\pi)\cos(\theta_{\mathrm{VB}})$ 
and 
$f_\mathrm{Dk_\perp} = {\boldsymbol V}_\perp \cdot 
{\boldsymbol k}_\perp /(2\pi) = (V k_\perp / 2\pi)\sin(\theta_{\mathrm{VB}}) 
({\boldsymbol V}_\perp \cdot \hat{\boldsymbol e}_{\perp_1}/V_\perp)$.  
The observed variation of the characteristic frequency associated with enhanced negative $\sigma_{\mathrm{m\perp_2\parallel}}$ is consistent with the $\theta_{\mathrm{VB}}$ dependence of $f_\mathrm{Dk_\perp}$ rather than that of $f_\mathrm{Dk_\parallel}$, as shown by the predicted frequency $f_\mathrm{Dk_\perp} = V_\perp /(2\pi\rho_p)$ for ion-scale waves with $k_\perp\rho_p = 1$ ($\rho_p \simeq 6.4$~km; grey curve in Figure~\ref{fig:2}(a)).  
Moreover, the intrinsic wave frequency $f_{\mathrm{pl}}$ depends on the wave vector (e.g., $f_{\mathrm{pl}} \propto k_\parallel$ for Alfv\'en-mode waves) and thus cannot account for the observed $\theta_{\mathrm{VB}}$ dependence.  
Taken together, these comparisons imply that waves with enhanced negative $\sigma_{\mathrm{m\perp_2\parallel}}$ exhibit highly oblique propagation ($k_\perp \gg k_\parallel$).

The distribution of $\sigma_{\mathrm{m\perp_1\perp_2}}$ in Figure \ref{fig:2}(b) shows a weak positive helicity signature within the region of enhanced negative $\sigma_{\mathrm{m\perp_2\parallel}}$ near $\theta_{\mathrm{VB}} \sim 90^\circ$ (Figure \ref{fig:2}(a)). These contrasting helicity signatures---enhanced negative $\sigma_{\mathrm{m\perp_2\parallel}}$ and weak positive $\sigma_{\mathrm{m\perp_1\perp_2}}$---are consistent with the polarization properties predicted for KAWs \citep{Gary1986JPlPh,Zhao2014ApJ,woodham2021dependence}. In addition, a band of weakly negative $\sigma_{\mathrm{m\perp_1\perp_2}}$ appears at $\theta_{\mathrm{VB}} \gtrsim 90^\circ$, indicative of left-hand wave signatures. The frequencies of these waves increase with $\theta_{\mathrm{VB}}$, consistent with Doppler-shifted frequencies $f = V \cos(\theta_{\mathrm{VB}} - 170^\circ) k/2\pi$, expected for quasi-parallel waves under the assumption of wave normal angles near $170^\circ$ and $\lambda_p k \approx 0.6$, where $\lambda_p$ is the proton inertial length. The origin of these left-hand waves remains uncertain and may be related either to residual coherent wave activity driven by plasma instabilities \citep{verniero2020parker,liu2023radial} or to turbulent cascades modified by a helicity barrier \citep{meyrand2021violation,squire2022high}. 
While weakly coherent waves, potentially ICWs or FMWs, remain in the turbulence dataset, their frequencies are largely confined below 2~Hz, near the transition from the MHD inertial range to the ion transition range (Figure~\ref{fig:2}(c)). This indicates that their contribution to the imbalance analysis at ion and sub-ion scales is likely limited.

In addition to the dependence of $\sigma_{\mathrm{m\perp_1\perp_2}}$ and $\sigma_{\mathrm{m\perp_2\parallel}}$ on $\theta_{\mathrm{VB}}$, the $P_B$ also exhibits a dependence on this angle \citep{duan2021anisotropy}. To illustrate the spectral properties, Figure \ref{fig:2}(c) shows the $P_B$, defined as $P_B(f,\theta_{\mathrm{VB}})=(2dt/N) \sum_i |{\boldsymbol W}(t_i,\theta_{\mathrm{VB}})|^2$, across four $\theta_{\mathrm{VB}}$ regimes: $75^\circ$–$90^\circ$, $90^\circ$–$105^\circ$, $105^\circ$–$120^\circ$, and $120^\circ$–$135^\circ$. Here, $dt$ represents the time resolution of the magnetic field data, and $N$ is the number of data points in each $\theta_{\mathrm{VB}}$ range. Assuming a power-law form for $P_B$, i.e., $P_B = C f^{\alpha}$, the spectral index $\alpha$ at each frequency $f$ is obtained by fitting $P_B$ over the range from $f/2$ to $2f$. 
The distribution of $\alpha$, shown in Figure \ref{fig:2}(c), reveals a pronounced steepening in the spectral slope between approximately 3 and 15 Hz, corresponding to the ion transition range. Based on the $P_B$ in the two angle regimes of  $75^\circ$–$90^\circ$ and $90^\circ$–$105^\circ$, the average spectral index within the ion transition range is approximately $-3.37 \pm 0.05$, compared to $-1.57 \pm 0.02$ in the inertial range (0.2–1 Hz) and $-2.63 \pm 0.01$ in the higher-frequency (20–100 Hz) kinetic-inertial range.

To explore the wavenumber corresponding to the transition range, Figure \ref{fig:2}(d) presents $P_B$ with $\theta_{\mathrm{VB}}$ restricted to the range $75^\circ$--$105^\circ$, calculated using  the expression \citep{2023ApJ...943L...8S}, 
$P_B (Lk^*,\theta_{\mathrm{VB}}) = (dt/\pi N) \sum_i [ V_\perp(t_i) |{\boldsymbol W}(t_i,\theta_{\mathrm{VB}})|^2 /L(t_i)]$, 
where $L$ presents either $\rho_p$ or $\lambda_p$ and $k^*=2\pi f/V_\perp$. 
Both $P_B(\rho_p k^* )$ and $P_B(\lambda_p k^* )$ display similar power-law behavior, with a spectral transition occurring at $\rho_p k^* \approx 0.4$--2.0 and $\lambda_p k^* \approx 0.7$--3.7. These results indicate that the transition from the inertial to transition range starts at a scale larger than the characteristic ion scales, consistent with the observations reported by \cite{bowen2020constraining}.


\section{Transition from imbalanced to balanced kinetic Alfv\'enic turbulence} 
 
\begin{figure}[!]
	\centering
	\includegraphics[width=3.5in]{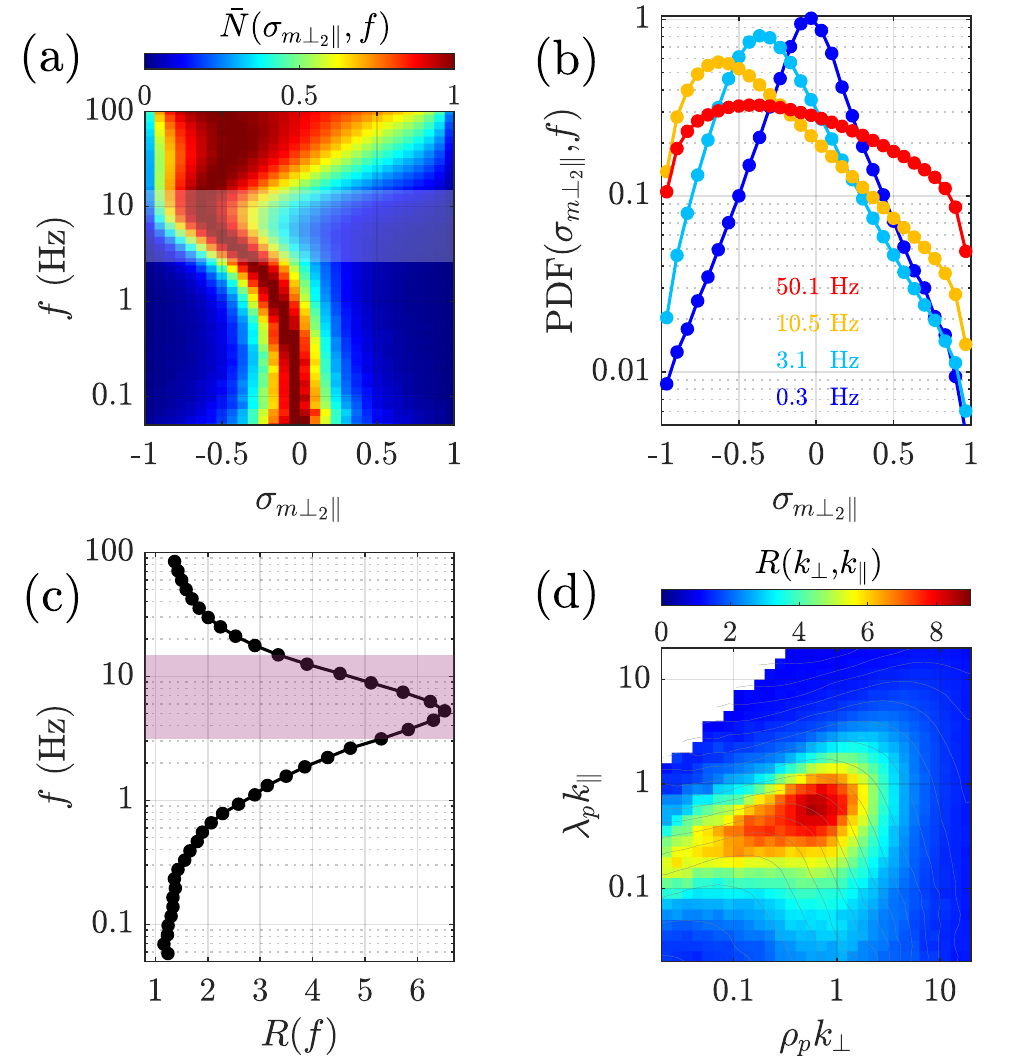}
	\caption{(a) The number of the data $N(\sigma_{\mathrm{m\perp_2\parallel}},f)$ normalized by the maximum $N(\sigma_{\mathrm{m\perp_2\parallel}},f)$ at each $f$, denoted by ${\bar N}(\sigma_{\mathrm{m\perp_2\parallel}},f)$. (b) The probability distribution function of $N(\sigma_{\mathrm{m\perp_2\parallel}},f)$, denoted by $\mathrm{PDF}$($\sigma_{\mathrm{m\perp_2\parallel}}$,$f$), at four typical frequencies: $f=$ 0.3, 3.1, 10.5, and 50.1 Hz. (c) The ratio $R(f)$ between the total data numbers with negative and positive $\sigma_{\mathrm{m\perp_2\parallel}}$,  $ R(f) = N_t(f,\sigma_{\mathrm{m\perp_2\parallel}} < 0)/N_t(f,\sigma_{\mathrm{m\perp_2\parallel}} > 0)$. (d) The ratio $R$ between the data numbers with negative and positive $\sigma_{\mathrm{m\perp_2\parallel}}$ in $k_\perp$--$k_\parallel$ space, where the grey curves are the contour lines of $P_B$. The data in (a)--(c) are limited to  $|\theta_\mathrm{VB}-90^\circ|<45^\circ$, and the data in (d) are limited to $|\sigma_{\mathrm{m\perp_2\parallel}}/\sigma_{\mathrm{m\parallel\perp_1}}|>2$; all data are subject to the condition  $|W_{\parallel}|^2/\left(|W_{\perp_1}|^2+|W_{\perp_2}|^2\right)<1$.
	}
    \label{fig:3}
\end{figure}

Because the sign of $\sigma_{\mathrm{m\perp_2\parallel}}$ serves as a diagnostic for the wave propagation direction, we can elucidate the degree of imbalance in kinetic Alfv\'enic turbulence by statistically analyzing the number of data points $N$ in the $\sigma_{\mathrm{m\perp_2\parallel}}–f$ space, as shown in Figure \ref{fig:3}.

Figure \ref{fig:3}(a) depicts the distribution of ${\bar N}(\sigma_{\mathrm{m\perp_2\parallel}},f)$, defined as $N(\sigma_{\mathrm{m\perp_2\parallel}},f)$ normalized by its maximum at each frequency $f$. Two primary features are evident: (1) as $f$ increases, the location of the maximum ${\bar N}=1$ rapidly shifts toward lower values of $\sigma_{\mathrm{m\perp_2\parallel}}$ within $\sim1$--$10$ Hz, primarily arising in the transition range; and (2) in the kinetic-inertial range, ${\bar N}$ exhibits a broader distribution of $\sigma_{\mathrm{m\perp_2\parallel}}$ compared to that in the transition range, suggesting a tendency toward balanced signatures at smaller scales.

Figure \ref{fig:3}(b) shows the probability distribution function (PDF) of $N(\sigma_{\mathrm{m\perp_2\parallel}}, f)$ at four representative frequencies: $0.3$, $3.1$, $10.5$, and $50.1~\mathrm{Hz}$. 
In the inertial range ($f=0.3~\mathrm{Hz}$), the PDF roughly follows a Gaussian distribution centered around zero, indicating that $\sigma_{\mathrm{m\perp_2\parallel}}$ behaves randomly. In the transition range ($f=3.1$ and $10.5~\mathrm{Hz}$), the PDFs are dominated by negative $\sigma_{\mathrm{m\perp_2\parallel}}$, reflecting a significant imbalance. At 50.1 $\mathrm{Hz}$, within the kinetic-inertial range, the PDF becomes broad and flat, indicating weakened imbalance  signature and a trend toward more balanced turbulence.

To quantify the degree of imbalance, we calculate the ratio $R$ of data numbers with negative to those with positive $\sigma_{\mathrm{m\perp_2\parallel}}$:
\[
R(f) = \frac{N_t(f,\sigma_{\mathrm{m\perp_2\parallel}} < 0)}{N_t(f,\sigma_{\mathrm{m\perp_2\parallel}} > 0)},
\]
as illustrated in Figure \ref{fig:3}(c). The value of $R$ increases with frequency, peaks ($R\sim 6.5$) near $f \sim 5\,\mathrm{Hz}$, and then decreases toward unity at higher frequencies. Since the transition range spans $f=3-15$ Hz, the position of this maximum roughly coincides with the onset of the transition range. The corresponding frequency $f \sim 5\,\mathrm{Hz}$ translates to spatial scales of $\rho_p k \sim 0.6$ and $\lambda_p k \sim 1.3$, which lie near the ion characteristic scale.

Figure \ref{fig:3}(d) presents the ratio $R$ between the data numbers with negative and positive $\sigma_{\mathrm{m\perp_2\parallel}}$ in the $k_\perp$--$k_\parallel$ space. Using the angle $\phi$ between ${\boldsymbol k}_\perp$ and ${\boldsymbol V}_\perp$, $k_\perp$ is determined via $k_\perp = 2\pi f \mathrm{sin}(\theta_\mathrm{VB}) /(V \mathrm{cos} (\phi))$ (see the sketch in Figure \ref{fig:a1}), while $k_\parallel$ is calculated as $k_\parallel=2\pi f |\mathrm{cos}(\theta_\mathrm{VB})|/V$. At scales with $\rho_pk_\perp \lesssim 0.2$, the maximum $R$ occurs at $\lambda_pk_\parallel \sim 0.3$. As $\rho_pk_\perp$ increases beyond 0.2, the region of maximum $R$ closely follows the core of the contour lines of $P_B$ (the grey curves in Figure \ref{fig:3}(d)) in the wave number space. At sub-ion scales with $\rho_p k_\perp \gtrsim 10$, the degree of imbalance tends to diminish, consistent with a trend toward more balanced characteristics at smaller scales.


\section{Conclusions and Discussion}

This Letter investigates kinetic Alfv\'enic turbulence in the near-Sun solar wind, utilizing magnetic field measurements from the PSP. To effectively identify wave modes, we introduce a magnetic helicity method that employs field-aligned coordinates defined by the background magnetic field and the wave vector. Our primary focus is on the reduced magnetic helicity $\sigma_{\mathrm{m\perp_2\parallel}}$, which quantifies the coherence between fluctuations in the parallel magnetic field and principal perpendicular magnetic field fluctuations. This methodology successfully identifies signatures of KAWs. Our analysis reveals the prevalence of KAWs characterized by negative values of $\sigma_{\mathrm{m\perp_2\parallel}}$, particularly when the observed wave frequency exceeds the proton cyclotron frequency.

Our examination of the $\mathrm{PDF}(\sigma_\mathrm{\perp_2\parallel}, f)$ and the ratio $R(f)$ reveals two distinct trends as wave frequency increases. First, the $\mathrm{PDF}(\sigma_\mathrm{\perp_2\parallel}, f)$ broadens with frequency. Second, $R(f)$ initially rises, reaches a peak, and then decreases. These trends reflect the scale-dependent evolution of KAW propagation: an initial dominance of outward-propagating waves cascading to smaller scales, followed by a relative increase in inward-propagating waves, leading to a more balanced state at smaller scales. This scenario is further supported by the ratio of outward to inward waves in the two-dimensional wavenumber space. This ratio distribution also reveals a notable feature of ion-scale turbulence: the maximum ratio closely aligns with the cascade direction inferred from the $P_B$ distribution in wavenumber space.

These findings prompt two critical questions regarding our understanding of solar wind turbulence at and below the ion scale.

The first question addresses why the imbalance increases toward the transition range. 
Our observations of imbalance development are consistent with the helicity barrier model \citep{meyrand2021violation, squire2022high}. The conservation of generalized helicity selectively suppresses the cascade of outward-propagating waves near and beyond the ion scale, leading to a pile-up of the waves at scales approaching $\rho_p k_\perp \sim 1$ \citep{squire2022high}. This pile-up and associated imbalance increase are evident in our observations. Moreover, as shown by the $P_B$ distribution in wavenumber space, the turbulent cascade preferentially channels energy toward $\lambda_p k_\parallel \sim 1$ and $\rho_p k_\perp \sim 1$, consistent with turbulent behavior predicted by numerical simulations of imbalanced Alfv\'enic turbulence influenced by the helicity barrier \citep{Squire2023ApJ, Zhang2025ApJ}. Additionally, although previous studies have suggested that plasma instabilities can generate KAWs in the solar wind \citep{Alexandrova2013SSRv, Hellinger2015ApJ}, direct observational evidence remains limited, and their potential contribution as a complementary source of KAWs near the ion scale warrants further investigation.

The second question involves the evolution of turbulence within the kinetic-inertial range. Our observations reveal that turbulence in this range remains weakly imbalanced down to sub-ion scales at $\rho_p k_\perp \sim 10$. This residual imbalance may indicate that energy transfer at kinetic scales is influenced by nonlocal interactions, such as couplings with larger-scale fluctuations or nonlocal spectral transfer processes \citep{Voitenko2005PhRvL,Lin2012PhRvL}. Moreover, differential damping rates for outward- and inward-propagating KAWs in the near-Sun solar wind, potentially affected by the presence of ion beam populations, may contribute to sustaining the weak imbalance. Turbulence in the kinetic-inertial range may also be influenced by additional processes, including KAW generation by plasma instabilities and the presence of coherent structures \citep{Vinogradov2024ApJ}. Our observations further show that the maximum imbalance tends to align with the cone in the $P_B$ distribution in wavenumber space at $\rho_p k_\perp \gtrsim 1$, consistent with predictions from numerical simulations \citep{squire2022high, Squire2023ApJ, Zhang2025ApJ}. This alignment may indicate a preferential cascade pathway at sub-ion scales, although the physical mechanisms governing this behavior remain to be fully understood.

In addition, the time interval analyzed contains switchbacks spanning a range of time scales, characterized by changes (including reversals) of the radial
magnetic field direction \citep{Dudok2020ApJS}. We therefore examine the imbalance distribution during magnetic field reversal intervals, as shown in Appendix~C. We observe the same transition from imbalanced to balanced turbulence within the reversal regions. This indicates that such field reversals do not change the wave propagation direction relative to the local background magnetic field, as illustrated by the schematic in Appendix~C. However, magnetic field reversals can change the apparent dominance of outward- versus inward-propagating Alfv\'enic fluctuations in the spacecraft frame, as reflected by the sign change of the cross helicity \citep{McManus2020ApJS}. Moreover, the reduced cross helicity within switchbacks \citep{Bourouaine2020ApJ} may lead to an imbalance evolution that quantitatively differs from that in the surrounding solar wind, since high cross helicity can induce a steeper transition range than in a low cross helicity environment \citep{McIntyre2025PhRvX}. A systematic comparison of turbulence imbalance inside switchbacks and in the ambient solar wind will be conducted in a future study.

In conclusion, the observations presented in this Letter emphasize the need for refined theoretical frameworks to accurately model the solar wind turbulence at kinetic scales.

\begin{acknowledgments}

This study is supported by the Strategic Priority Research Program of the Chinese Academy of Sciences (Grant No. XDB0560000) and the National Key R\&D Program of China 2022YFF0503000 (2022YFF0503003). Z.J.S. thanks for the support from NSFC 42374196. TDdW acknowledges support from CNES. The FIELDS and SWEAP instruments were designed and built under NASA Contract No. NNN06AA01C. 

\end{acknowledgments}

\appendix

\section{Three magnetic helicity methods}

In this section, we introduce three methods for calculating the reduced magnetic helicity in solar wind turbulence: the RTN-coordinate method (Subsection A.1), the ${\boldsymbol B}_0$\&${\boldsymbol V}_0$--FAC method (Subsection A.2), and the ${\boldsymbol B}_0$\&${\boldsymbol k}$--FAC method developed in this paper (Subsection A.3). We also compare the results from the two FAC methods to demonstrate the advantages of our method (Subsection A.4).

\subsection{Method I: RTN-coordinate method}

Using the two magnetic field components $B_T$ and $B_N$ in radial--tangential--normal (RTN) coordinates, \cite{Matthaeus1982JGR} first employed magnetic helicity to analyze solar wind turbulence. The reduced magnetic helicity is defined as $\sigma_{\mathrm{mTN}}(f) = 2\,\mathrm{Im}(W_T W^{*}_N)/|{\boldsymbol W}|^2$, where $W_T$, $W_N$, and $|{\boldsymbol W}|$ represent the complex amplitudes of $B_T$, $B_N$, and the magnitude of the complex wave amplitude at frequency $f$, respectively.

The distribution of $\sigma_{\mathrm{mTN}}$ as a function of the angle $\theta_\mathrm{VB}$ is widely used to diagnose ICWs and KAWs in the solar wind
\citep{he2011possible,He2012ApJ,podesta2011magnetic,Klein2014ApJ,huang2020kinetic}, where $\theta_\mathrm{VB}$ is the angle between the solar wind magnetic field ${\boldsymbol B}$ and bulk velocity ${\boldsymbol V}$. Positive (negative) values of $\sigma_{\mathrm{mTN}}$ in the outward (inward) magnetic-field sectors, particularly near $\theta_{\mathrm{VB}} \sim 90^\circ$, are characteristic signatures of KAWs \citep{he2011possible,He2012ApJ,podesta2011magnetic,Klein2014ApJ,huang2020kinetic}. By contrast, finite $\sigma_{\mathrm{mTN}}$ values near $\theta_{\mathrm{VB}} \rightarrow 0^\circ$ or $180^\circ$ indicate the presence of quasi-parallel coherent ion-scale waves.

\subsection{Method II: ${\bf B}_0$\&${\bf V}_0$-FAC method}

To separate the signatures of highly oblique KAWs from quasi-parallel coherent ion-scale waves, \cite{woodham2021dependence} developed an identification method based on magnetic helicity defined in a FAC system, constructed using the background magnetic field ${\boldsymbol B}_0$ and solar wind velocity
${\boldsymbol V}_0$. The three unit axes are defined as:
$\hat{\boldsymbol e}_{\perp_1^{'}} \equiv \hat{\boldsymbol e}_{\perp_2^{'}} \times \hat{\boldsymbol e}_{\parallel}$, $\hat{\boldsymbol e}_{\perp_2^{'}} \equiv ({\boldsymbol B}_0 \times {\boldsymbol V}_0)/(|{\boldsymbol B}_0||{\boldsymbol V}_0|)$, and $\hat{\boldsymbol e}_{\parallel} \equiv {\boldsymbol B}_0 / |{\boldsymbol B}_0|$. The reduced magnetic helicity components are then defined as:
\begin{eqnarray}
\sigma_{\mathrm{m\perp_1^{'}\perp_2^{'}}}    &=& -2\, \mathrm{Im}(W_{\perp_1^{'}} W^{*}_{\perp_2^{'}})/|{\boldsymbol W}|^2, \\
\sigma_{\mathrm{m\perp_2^{'}\parallel}}  &=& -2\, \mathrm{Im}(W_{\perp_2^{'}} W^{*}_{\parallel})/|{\boldsymbol W}|^2, \\
\sigma_{\mathrm{m\parallel\perp_1^{'}}} &=& -2\, \mathrm{Im}(W_{\parallel} W^{*}_{\perp_1^{'}})/|{\boldsymbol W}|^2,
\end{eqnarray}
where $|{\boldsymbol W}|^2 = |W_{\perp_1^{'}}|^2 + |W_{\perp_2^{'}}|^2 + |W_{\parallel}|^2$.

Using $\sigma_{\mathrm{m\perp_1^{'}\perp_2^{'}}}$ and $\sigma_{\mathrm{m\perp_2^{'}\parallel}}$, quasi-parallel coherent ion-scale waves and KAWs can be separately diagnosed \citep{woodham2021dependence}.  Specifically, $\sigma_{\mathrm{m\perp_2^{'}\parallel}}$ is suggested as a reliable indicator for identifying KAWs. In addition, $\sigma_{\mathrm{m\parallel\perp_1^{'}}}$ also captures the characteristic signatures of KAWs.

\subsection{Method III: ${\bf B}_0$\&${\bf k}$-FAC method}

Because KAWs are highly elliptical or even linearly polarized in the perpendicular direction \citep{Hollweg1999JGR,Zhao2014ApJ,Zhao2022ApJ}, their major perpendicular magnetic perturbations do not always coincide with either $W_{\perp_1^{'}}$ or $W_{\perp_2^{'}}$. Consequently, both $\sigma_{\mathrm{m\perp_1^{'}\parallel}}$ and $\sigma_{\mathrm{m\perp_2^{'}\parallel}}$ tend to underestimate the coherence between the parallel and the major perpendicular magnetic field perturbations of KAWs.

To address this limitation in Method II, we develop a new wave diagnostic method based on a FAC system defined by the wave vector ${\boldsymbol k}$ and ${\boldsymbol B}_0$. In this system, the three axes are given by $\hat{\boldsymbol e}_{\mathrm{\perp_1}} \equiv \hat{\boldsymbol e}_{\mathrm{\perp_2}} \times \hat{\boldsymbol e}_{\parallel}$, $\hat{\boldsymbol e}_{\mathrm{\perp_2}} \equiv ({\boldsymbol B}_0 \times {\boldsymbol k}_\perp)/(|{\boldsymbol B}_0||{\boldsymbol k}_\perp|)$, and $\hat{\boldsymbol e}_{\parallel}$. The major magnetic perturbation of KAWs, $W_{\perp_2}$, resides along the $\hat{\boldsymbol e}_{\mathrm{\perp_2}}$ direction. Therefore, the coherence between $W_{\perp_2}$ and $W_{\parallel}$ serves as a reliable parameter for identifying KAWs.

\begin{figure}[t]
\centering
\includegraphics[width=4.5in]{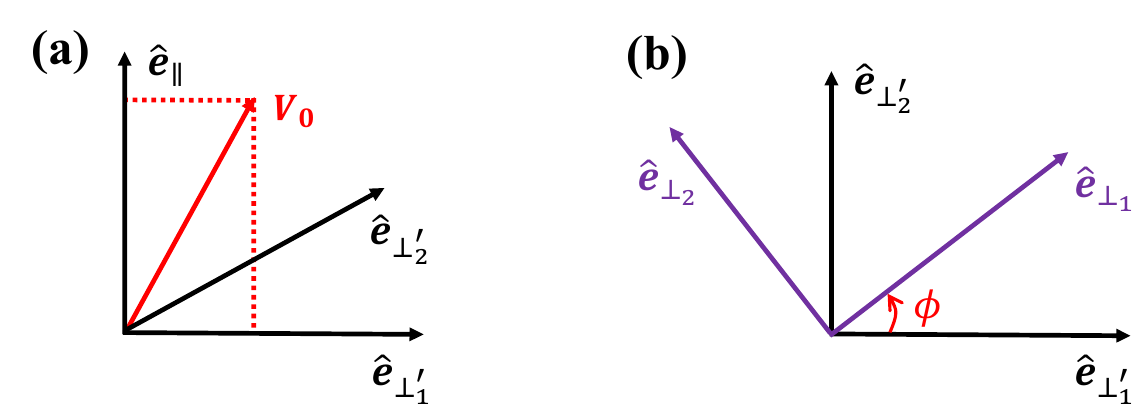}
\caption{ 
(a) The field-aligned coordinate (FAC) system defined by the background magnetic field ${\boldsymbol B_0}$ and solar wind velocity ${\boldsymbol V_0}$, where $\hat{\boldsymbol e}_{\parallel} \equiv {\boldsymbol B_0} / {|{\boldsymbol B_0}|}$, $\hat{\boldsymbol e}_{\perp_1^{'}} \equiv \hat{\boldsymbol e}_{\perp_2^{'}} \times  \hat{\boldsymbol e}_{\parallel}$, and $\hat{\boldsymbol e}_{\perp_2^{'}} \equiv {\boldsymbol B_0}\times{\boldsymbol V_0}/(|{\boldsymbol B_0}||{\boldsymbol V_0}|)$. (b) The rotation from $(\hat{\boldsymbol e}_{\perp_1^{'}}, \hat{\boldsymbol e}_{\perp_2^{'}})$ into $(\hat{\boldsymbol e}_{\perp_1}, \hat{\boldsymbol e}_{\perp_2})$, where $\phi$ represents the rotation angle between the two sets of coordinates.
}
\label{fig:a1}
\end{figure}

In the super-Alfv\'enic solar wind, the variability of the perpendicular wave vector ${\boldsymbol k}_\perp$ in the spacecraft frame exhibits the following characteristic property. Because low-frequency MHD- and kinetic-scale Alfv\'en mode waves are convected by the super-Alfv\'enic flow, the observed wave propagation in the plane defined by the magnetic field and flow directions is dominated by advection, while the wave propagation perpendicular to this plane is unaffected. As a result, the component of ${\boldsymbol k}_\perp$ along the $\hat{\boldsymbol e}_{\perp^{'}_1}$ direction remains fixed in sign, whereas its component along the $\hat{\boldsymbol e}_{\perp^{'}_2}$ direction can be either parallel or anti-parallel to this axis. This variability implies that the angle $\phi$ between ${\boldsymbol k}_\perp$ and $\hat{\boldsymbol e}_{\perp^{'}_1}$ can range from $-\pi/2$ to $\pi/2$ (see Figure~\ref{fig:a1}).

Given the range of $\phi$, we determine ${\boldsymbol k}_\perp$ based on the wave polarization properties in the $\hat{\boldsymbol e}_{\perp_1}$–$\hat{\boldsymbol e}_{\perp_2}$ plane. Using the relations $W_{\perp_1} = W_{\perp_1^{'}} \cos(\phi) + W_{\perp_2^{'}} \sin(\phi)$ and $W_{\perp_2} = -W_{\perp_1^{'}} \sin(\phi) + W_{\perp_2^{'}} \cos(\phi)$ in this plane, the direction of ${\boldsymbol k}_\perp$ is identified by finding the value of $\phi$ that maximizes $|W_{\perp_2}|$ as $\phi$ varies from $-\pi/2$ to $\pi/2$. Simultaneously, we obtain the two perpendicular magnetic field perturbation components in the $(\hat{\boldsymbol e}_{\perp_1}, \hat{\boldsymbol e}_{\perp_2}, \hat{\boldsymbol e}_\parallel)$ coordinate system.

This approach may not accurately characterize the perpendicular wave vector for quasi-parallel waves, as the two perpendicular magnetic field components are approximately equal in magnitude. However, their polarization characteristics (left-hand or right-hand) remain consistent across both sets of FACs.

\subsection{The comparison for the magnetic helicity distributions from the two FAC methods}

\begin{figure*}[t]
\centering
\includegraphics[width=6.5in]{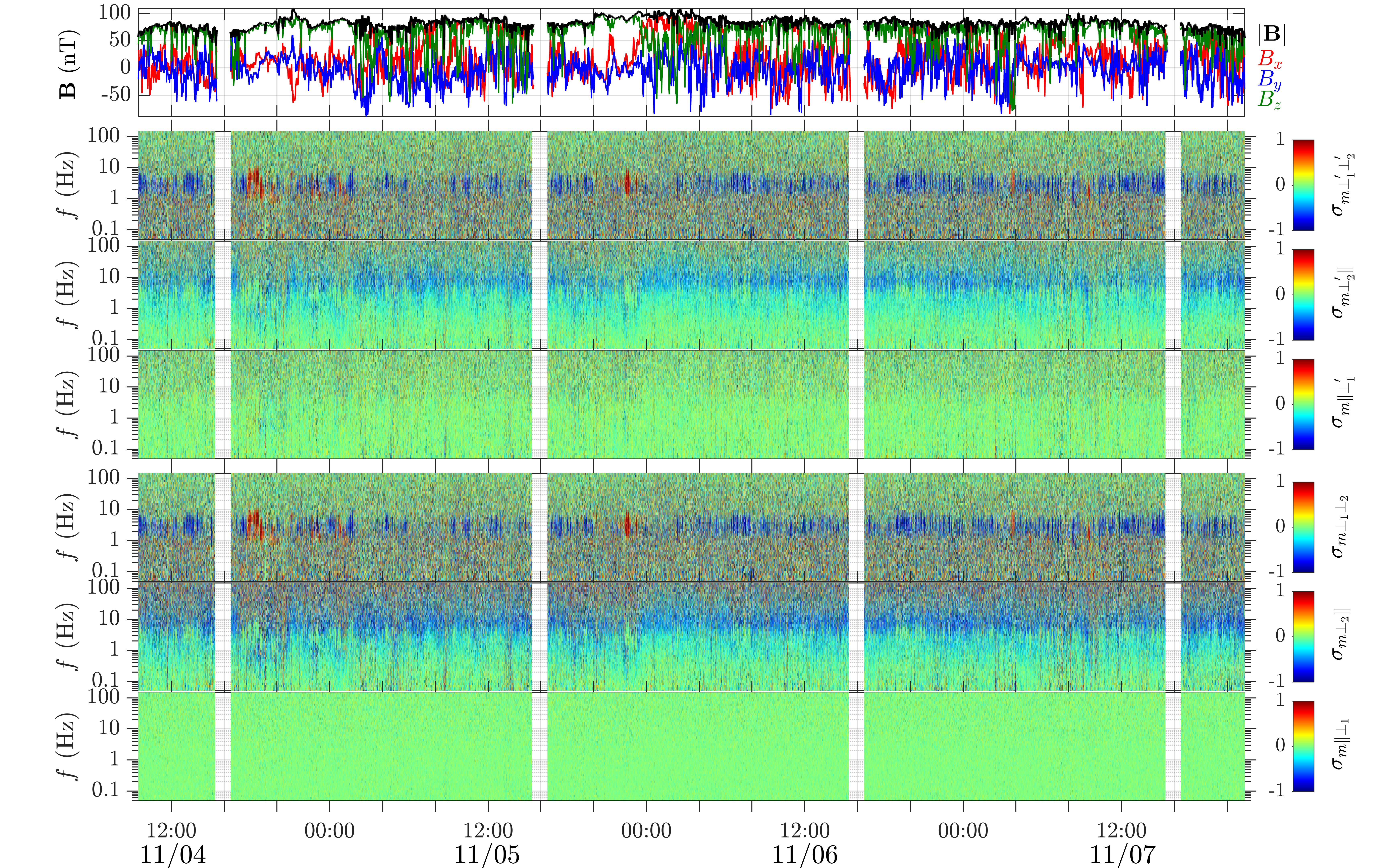}
\caption{ 
Overview of the reduced magnetic helicity distributions obtained from the two FAC methods. Top panel: The magnitude and the three components of the magnetic field in the spacecraft frame. Second to fourth panels: Distributions of $\sigma_\mathrm{m\perp^{'}_1 \perp^{'}_2}$, $\sigma_\mathrm{m\perp_2^{'} \parallel}$, and $\sigma_\mathrm{m\parallel \perp_1^{'}}$ from Method II. Fifth to seventh panels: Distributions of $\sigma_\mathrm{m\perp_1 \perp_2}$, $\sigma_\mathrm{m\perp_2 \parallel}$, and $\sigma_\mathrm{m\parallel \perp_1}$ from Method III.
}
\label{fig:a2}
\end{figure*}

\begin{figure*}[h]
\centering
\includegraphics[width=6.5in]{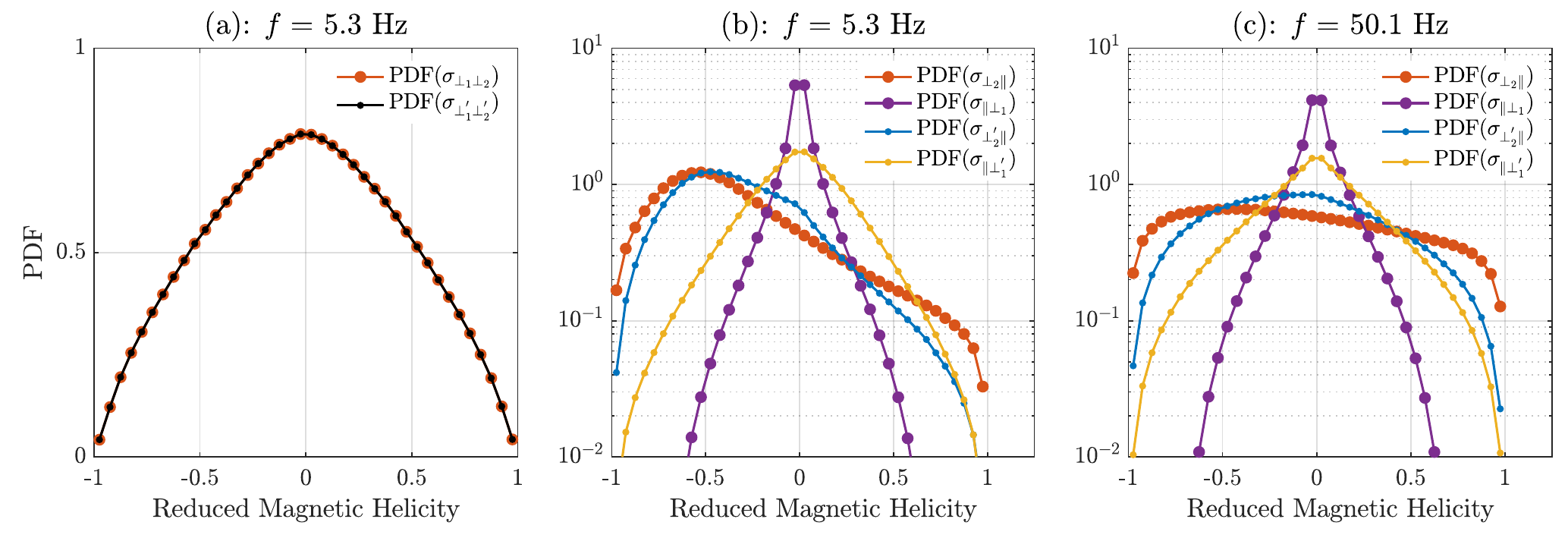}
\caption{ 
Comparison of the two FAC methods based on the PDFs of the reduced magnetic helicity. (a) PDFs of $\sigma_\mathrm{m\perp_1 \perp_2}$ (red) and $\sigma_\mathrm{m\perp^{'}_1 \perp^{'}_2}$ (black) at $f = 5.3$ Hz for all data. (b) PDFs of $\sigma_\mathrm{m\perp_2 \parallel}$ (red), $\sigma_\mathrm{m\perp_2^{'} \parallel}$ (blue), $\sigma_\mathrm{m\parallel \perp_1}$ (purple), and $\sigma_\mathrm{m\parallel \perp_1^{'}}$ (yellow) at $f = 5.3$ Hz for the identified turbulent waves. (c) The same four PDFs at $f = 50.1$ Hz for the identified turbulent waves.
}
\label{fig:a3}
\end{figure*}

Here we compare the reduced magnetic helicity distributions obtained from the two FAC methods.

To calculate the magnetic field perturbations at each frequency and time, we apply the Morlet wavelet transform to the time-series magnetic field data merged from the fluxgate and search coil magnetometers on PSP/FIELDS \citep{bowen2020merged}. Narrowband noise induced by PSP’s reaction wheels \citep{bowen2020ion} is removed using short-time Fourier transform techniques. Because the fast Fourier transform in wavelet analysis assumes waveforms in the form $e^{i\omega t}$, which differs from the conventional form $e^{-i\omega t}$, we apply a conjugate transformation to the complex wavelet amplitude so that it follows  the standard definition of wave evolution as $e^{-i\omega t}$. In addition, we average the wavelet amplitudes over eight periods to compute the coherence at each scale before calculating the magnetic helicity.

We construct the ${\boldsymbol B}_0 \& {\boldsymbol V}_0$ FAC system $(\hat{\boldsymbol e}_{\perp^{\prime}_1}, \hat{\boldsymbol e}_{\perp^{\prime}_2}, \hat{\boldsymbol e}_{\parallel})$, in which the local background magnetic field ${\boldsymbol B}_0$ and solar wind velocity ${\boldsymbol V}_0$ are determined following \citet{podesta2011magnetic}:
\begin{equation}
 {\boldsymbol B}_0(t_i) = \frac{1}{A} \sum_{t_j=t_i-3s}^{t_i+3s} {\boldsymbol B}(t_j)
 \exp\!\left[ - \frac{(t_j - t_i)^2}{2s^2} \right],
 \label{B0}
\end{equation}
\begin{equation}
 {\boldsymbol V}_0(t_i) = \frac{1}{A} \sum_{t_j=t_i-3s}^{t_i+3s} {\boldsymbol V}(t_j)
 \exp\!\left[ - \frac{(t_j - t_i)^2}{2s^2} \right],
 \label{V0}
\end{equation}
where
$A = \sum_{t_j=t_i-3s}^{t_i+3s} \mathrm{exp} \left[ - \frac{(t_j - t_i)^2}{2s^2} \right] $ is the normalization factor, and the wavelet scale $s$ ranges from $s\simeq 0.007$ s to $s=20$ s, with a total of 47 points in logarithmic space. Given the distinct temporal resolutions of the magnetic field and plasma measurements, ${\boldsymbol V}_0$ is resampled to match the temporal resolution of the magnetic field data. In addition to defining the ${\boldsymbol B}_0 \& {\boldsymbol V}_0$ FAC system, the scale-dependent backgrounds ${\boldsymbol B}_0$ and ${\boldsymbol V}_0$ are used to determine the angle between the solar wind velocity and the magnetic
field, $\theta_\mathrm{VB} = {\boldsymbol V}_0 \cdot {\boldsymbol B}_0 /\left( |{\boldsymbol V}_0|\,|{\boldsymbol B}_0| \right)$, and the angle between the instantaneous magnetic field and the background field, $\theta_\mathrm{BB_0} = {\boldsymbol B} \cdot {\boldsymbol B}_0 /\left( |{\boldsymbol B}|\,|{\boldsymbol B}_0| \right)$. Using the scale-dependent backgrounds ${\boldsymbol B}_0$ and ${\boldsymbol V}_0$ significantly improves the identification of the wave direction in the presence of rapid variations of the magnetic field and solar wind velocity, as expected during intervals with switchbacks.

The wavelet-transformed magnetic field perturbations are first reconstructed in the ${\boldsymbol B}_0 \& {\boldsymbol V}_0$ FAC system, yielding the complex wavelet amplitude ${\boldsymbol W}(f,t)$ at frequency
$f = \left(w_0 + \sqrt{2 + w_0^2}\right)/{4\pi s}$ \citep{Torrence1998BAMS}. We then apply the ${\boldsymbol B}_0 \& {\boldsymbol k}$-FAC method to obtain the magnetic field perturbations in the FAC system
$(\hat{\boldsymbol e}_{\perp_1}, \hat{\boldsymbol e}_{\perp_2}, \hat{\boldsymbol e}_{\parallel})$.

Using these FAC magnetic field perturbations, the distributions of the three reduced magnetic helicity components are shown in Figure~\ref{fig:a2}. We find that the distributions of
$\sigma_\mathrm{m\perp^{'}_1 \perp^{'}_2}$ and $\sigma_\mathrm{m\perp_1 \perp_2}$ are nearly identical, whereas the other four helicity components, which are relevant for diagnosing KAWs, differ significantly.

To further compare the results from the two FAC methods, the PDFs of the reduced magnetic helicity components are presented in Figure \ref{fig:a3}.

The PDFs of $\sigma_\mathrm{m\perp^{'}_1 \perp^{'}_2}$ and $\sigma_\mathrm{m\perp_1 \perp_2}$ at $f=5.3~\mathrm{Hz}$ for all data are shown in Figure \ref{fig:a3}(a). The two PDFs are consistent, indicating that both FAC methods play an equivalent role in quantifying quasi-parallel coherent waves.

To highlight the differences in quantifying KAWs, we select turbulent wave data using the method described in the next section. The four PDFs at $f = 5.3~\mathrm{Hz}$ and $50.1~\mathrm{Hz}$ differ substantially, as shown in Figures \ref{fig:a3}(b) and (c). In particular, the PDFs of $\sigma_{\mathrm{m}\perp_2\parallel}$ are significantly higher than those of $\sigma_{\mathrm{m}\perp^{\prime}_2\parallel}$ when their absolute values exceed $0.5$. Moreover, the PDFs of $\sigma_{\mathrm{m}\parallel\perp_1}$ peak around $\sigma_{\mathrm{m}\parallel\perp_1} = 0$, in agreement with theoretical predictions.

These comparisons between the two FAC methods demonstrate the effectiveness of our improvement in quantifying KAWs in solar wind turbulence.

\section{Identification of turbulent and coherent waves}

\begin{figure}[t]
\centering
\includegraphics[width=6.5in]{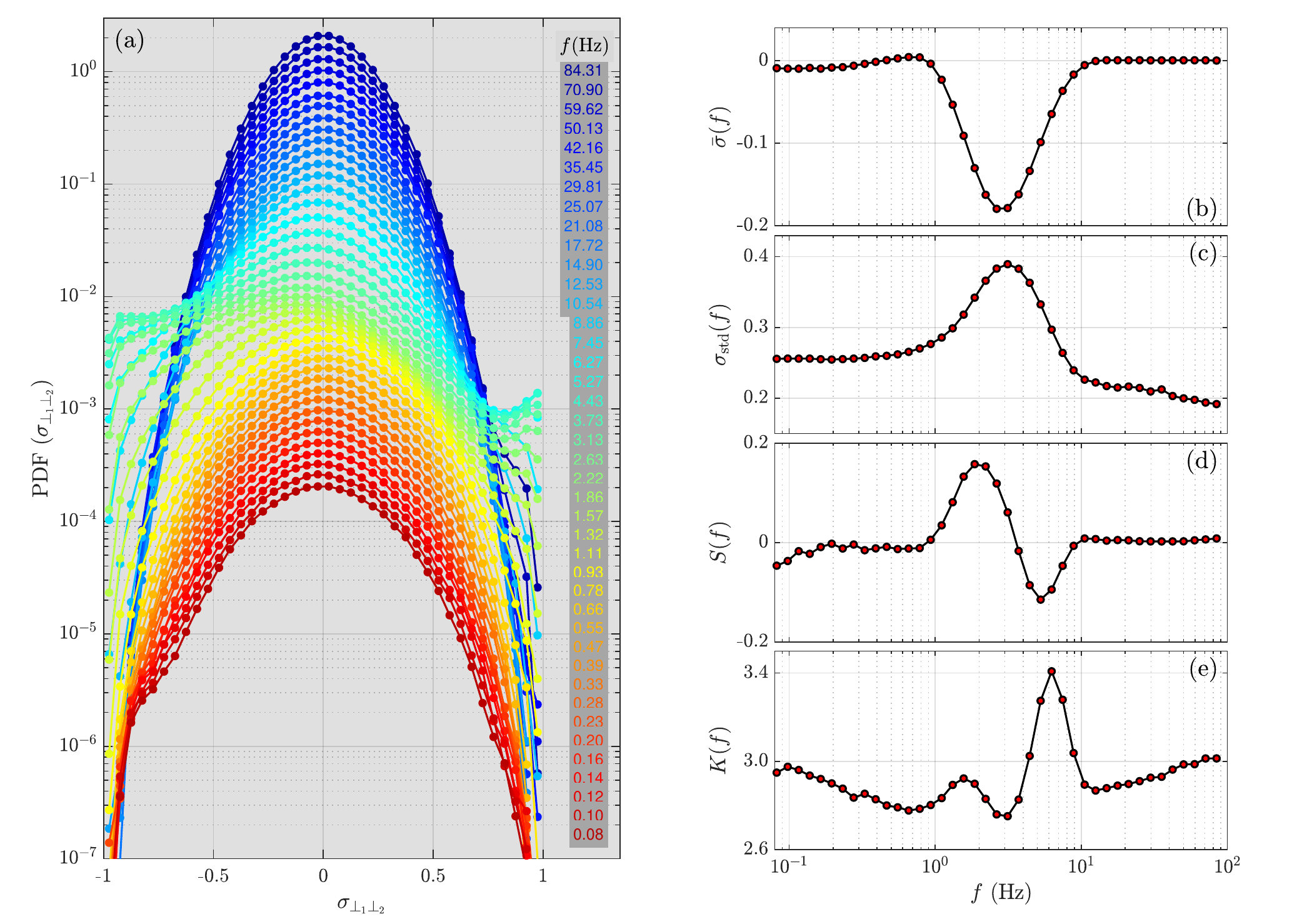}
\caption{ 
The PDFs of $\sigma_\mathrm{m\perp_1\perp_2}$ and their characteristic statistical parameters for all waves. (a) The PDFs of $\sigma_\mathrm{m\perp_1\perp_2}$ at frequencies between 0.08 and 84.31 Hz, where each PDF is multiplied by a factor of $0.8^{-[4 \times \log_2(84.31/f)]}$ to improve visualization. (b) The mean value ${\bar \sigma}$, (c) the standard deviation $\sigma_\mathrm{std}$, (d) the skewness $S$, and (e) the kurtosis $K$ corresponding to each PDF.
}
\label{fig:a4}
\end{figure}

\begin{figure}[t]
\centering
\includegraphics[width=6.5in]{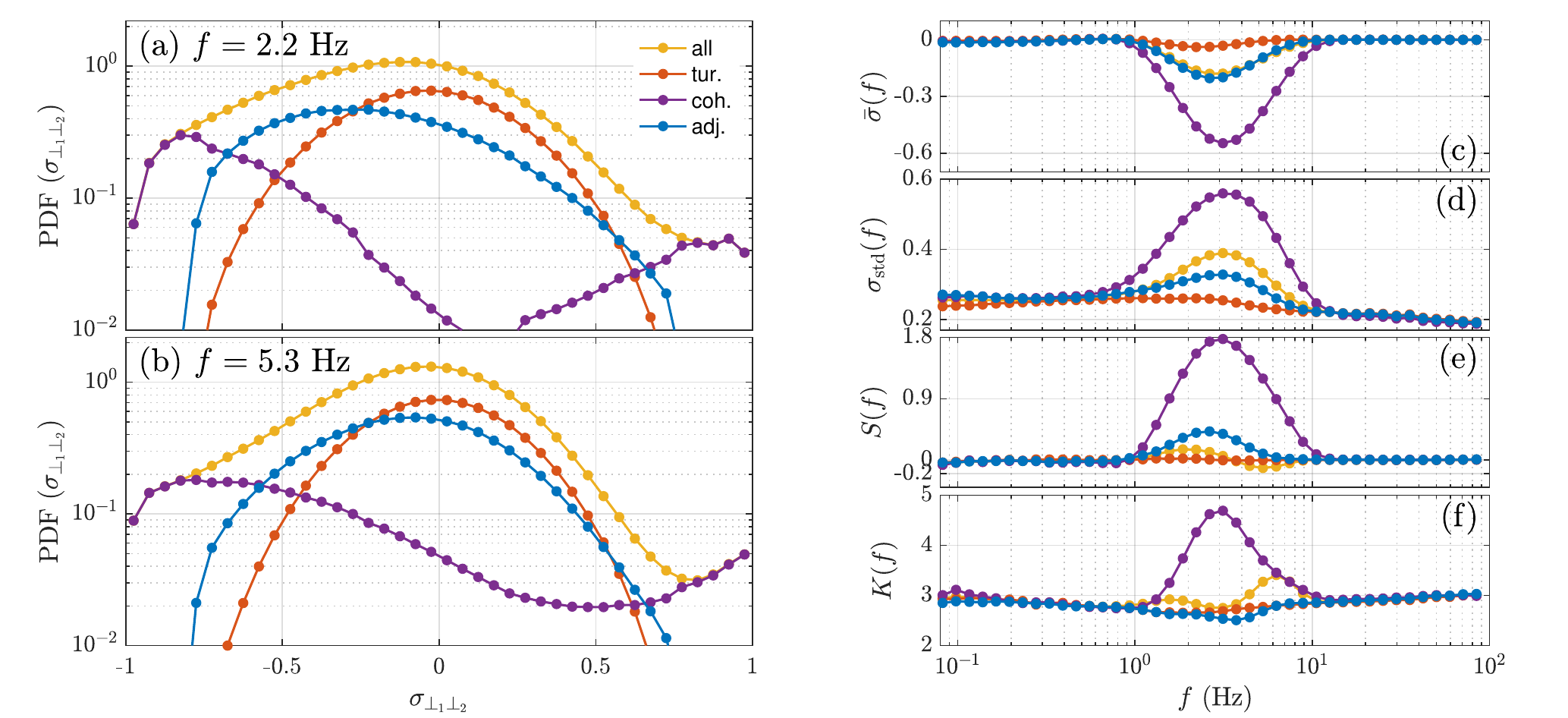}
\caption{ 
The PDFs of $\sigma_\mathrm{m\perp_1\perp_2}$ for the identified coherent (purple), adjacent (blue), and turbulent (red) waves, as well as for all waves (yellow): (a) at $f=2.2$ Hz and (b) at $f=5.3$ Hz. Panels (c)–(f) show the same parameters as those shown in Figure~\ref{fig:a4}, but for the identified coherent (purple), adjacent (blue), and turbulent (red) waves, along with all waves (yellow).
}
\label{fig:a5}
\end{figure}

\begin{figure}[t]
\centering
\includegraphics[width=7in]{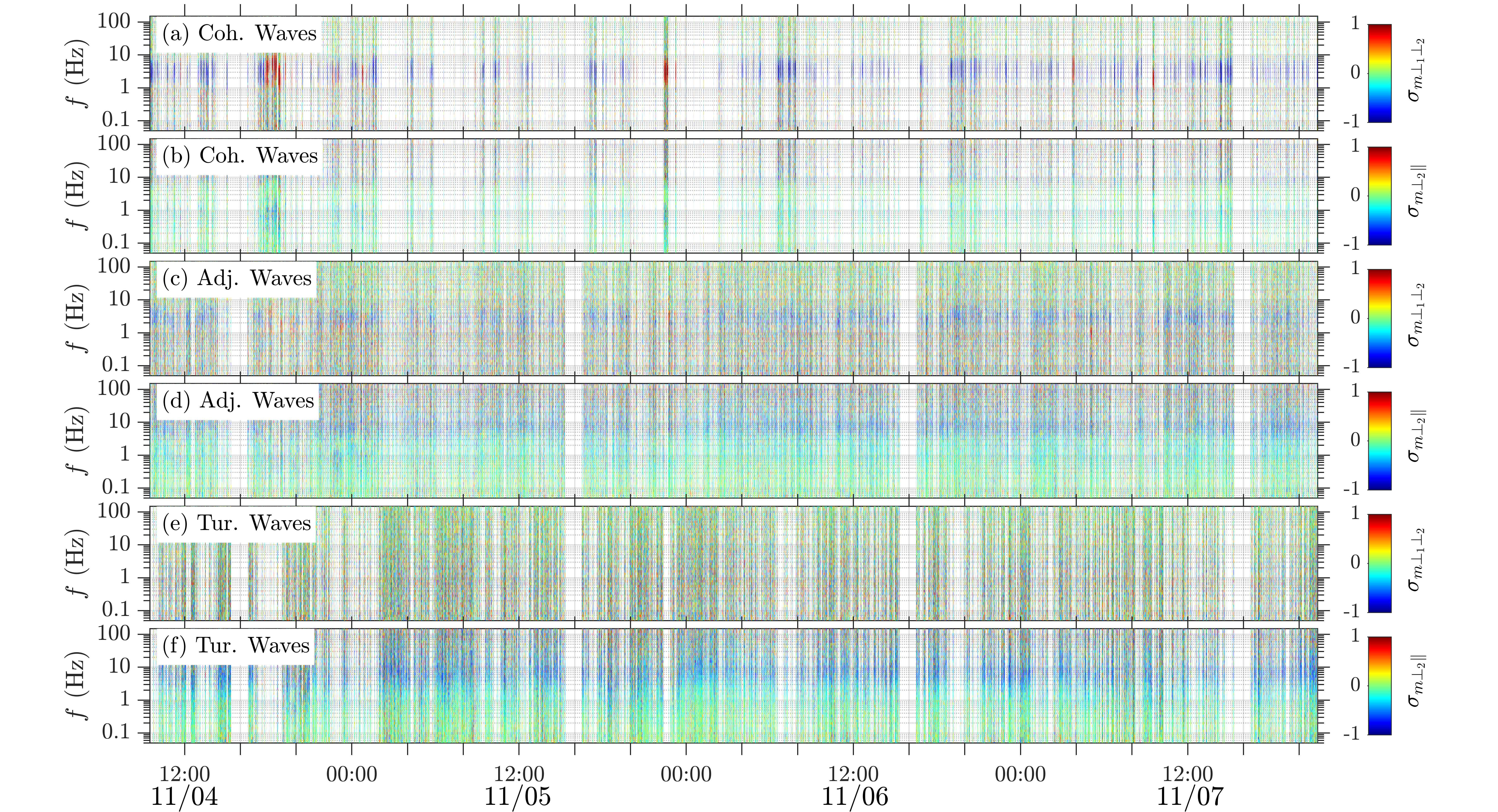}
\caption{ 
The distributions of the reduced magnetic helicities $\sigma_\mathrm{m\perp_1\perp_2}$ and $\sigma_\mathrm{m\perp_2\parallel}$ for coherent waves (a, b), adjacent waves (c, d), and turbulent waves (e, f).
}
\label{fig:a6}
\end{figure}

During the time interval of interest, both turbulent waves and quasi-parallel coherent waves are present, as shown in Figure~\ref{fig:1} and Figure~\ref{fig:a2}. To isolate the intervals dominated by turbulent waves, a statistical identification method based on the signature of $\sigma_\mathrm{m\perp_1 \perp_2}$ is developed for this purpose.

The primary motivation for our identification method is derived from the distributional properties of the PDFs of $\sigma_\mathrm{m\perp_1 \perp_2}$, as shown in Figure~\ref{fig:a4}. To better capture the statistical characteristics, the values of $\sigma_\mathrm{m\perp_1 \perp_2}$ are averaged over a smoothed frequency band between $f_L$ and $f_U$, where $f_U/f_L = 2$.

Furthermore, to quantitatively characterize these PDFs, we calculate their statistical moments, including the mean value $\bar{\sigma}=\sum_{i=1}^N \sigma_i/N$ in Figure~\ref{fig:a4}(b), the standard deviation $\sigma_{\mathrm{std}}=\sqrt{\sum_{i=1}^N \left(\sigma_i - \bar{\sigma}\right)^2/N}$ in Figure~\ref{fig:a4}(c), the skewness $S= \frac{1}{N \, \sigma_{\mathrm{std}}^3} \sum_{i=1}^N \left(\sigma_i - \bar{\sigma}\right)^3$ in Figure~\ref{fig:a4}(d), and the kurtosis $K=\frac{1}{N \, \sigma_{\mathrm{std}}^4} \sum_{i=1}^N \left(\sigma_i - \bar{\sigma}\right)^4$ in Figure~\ref{fig:a4}(e), where $N$ denotes the number of data points. Among these moments, the skewness quantifies the asymmetry of the distribution, with positive (negative) values indicating a longer tail to the right (left) of the mean, while the kurtosis describes the peakedness and tail heaviness of the distribution.

The PDFs of $\sigma_\mathrm{m\perp_1\perp_2}$ in Figure \ref{fig:a4} exhibit approximately Gaussian shapes in two distinct frequency ranges: $f = 0.08$--$0.8$ Hz and $10$--$100$ Hz. In both ranges, the skewness $S$ is close to zero, as expected for an ideal Gaussian distribution. The kurtosis is approximately 2.8--3.0 in the low-frequency range ($f = 0.08$--$0.8$ Hz), slightly below the value of 3 expected for a Gaussian distribution, while in the high-frequency range ($f = 10$--$100$ Hz) the kurtosis gradually increases toward 3 as the frequency increases.

However, a distinct departure from Gaussian behavior is observed in the intermediate-frequency range $f = 0.8$–$10$ Hz. Specifically, the distributions exhibit pronounced heavy tails at large values of $|\sigma_\mathrm{m\perp_1\perp_2}|$, predominantly attributable to  the presence of quasi-parallel coherent waves such as ICWs and FMWs. If these coherent wave events are excluded from the dataset within this frequency range, one would expect the resulting PDFs to resemble those observed in the low and high frequency ranges, displaying Gaussian-like features consistent with stochastic fluctuations.

Based on this observation, we develop a systematic procedure to process the $\sigma_\mathrm{m\perp_1\perp_2}$ data and identify coherent wave events. First, we exclude data points with $|\sigma_\mathrm{m\perp_1\perp_2}|$ smaller than a threshold $\sigma_\mathrm{t}$, defined as $\bar{\sigma}_\mathrm{LR} \simeq 0.26$, which corresponds to the mean value of $\sigma_{\mathrm{std}}$ within the low-frequency range ($0.08$--$0.8$ Hz). This step effectively removes weak, incoherent fluctuations. Next, at each time step, we determine the widest continuous frequency band containing the remaining data and identify key wave parameters, including the lower and upper frequency boundaries ($f_\mathrm{LB}$ and $f_\mathrm{UB}$) and the frequency $f_\mathrm{max}$ at which the maximum $|\sigma_\mathrm{m\perp_1\perp_2}|$ occurs (denoted as $\sigma_\mathrm{max}$). Finally, to classify strongly coherent waves, we apply a helicity criterion of $|\sigma_m| > 3\bar{\sigma}_\mathrm{LR}$ and the frequency band criterion of $f_\mathrm{UB}/f_\mathrm{LB}>2$.

In the regions adjacent to the identified coherent waves, the helicity signatures of nearby waves can be affected by the coherent waves themselves due to the finite width of the wavelet window used in the wavelet transform. Moreover, waves within these adjacent regions may exhibit moderate coherence, either as a result of contamination from nearby coherent waves or because they themselves possess weak intrinsic coherence. To account for this effect, we define an exclusion region around each coherent wave event, spanning from $t_\mathrm{coh} - \Delta t$ to $t_\mathrm{coh} + \Delta t$, where $t_\mathrm{coh}$ denotes the time location of the coherent wave and $\Delta t = 1/f_\mathrm{max}$ represents the characteristic timescale associated with the edge effect at frequency $f_\mathrm{max}$. By excluding data within these intervals, we ensure that the remaining dataset for turbulent wave analysis is minimally affected by adjacent coherent wave signatures.

By applying the aforementioned procedures, we classify the dataset into coherent waves, waves in regions adjacent to coherent waves (termed adjacent waves), and turbulent waves. The resulting distribution of wave populations comprises 16\% coherent waves, 44\% adjacent waves, and 40\% turbulent waves.

The validation of the identified results are shown in Figures \ref{fig:a5} and \ref{fig:a6}.

Figure \ref{fig:a5} presents the statistical properties of the three identified types of waves. Figures \ref{fig:a5}(a) and (b) display the PDFs of the identified turbulent, coherent, and adjacent waves at two representative frequencies, 2.2 Hz and 5.3 Hz, within the intermediate frequency range. The PDFs of the coherent waves exhibit pronounced deviations from a  Gaussian distribution, with significant values of $\sigma_\mathrm{m\perp_1\perp_2}$. In contrast, the PDFs of turbulent waves closely follow a Gaussian-like shape. The PDFs for adjacent waves show intermediate behavior between these two populations. The corresponding statistical moments, shown in Figures \ref{fig:a5}(c)–(f), further support these observations. The turbulent waves maintain relatively stable and Gaussian-consistent moment values across the entire frequency range, while the coherent waves display pronounced enhancements in standard deviation, skewness, and kurtosis near their dominant frequencies. The adjacent wave data exhibit moderate departures from Gaussianity, reflecting the influence of nearby coherent wave structures.

Figure \ref{fig:a6} shows the distributions of the two reduced magnetic helicities for the identified categories of coherent, adjacent, and turbulent waves. Coherent waves (a and b) exhibit localized, intermittent high-$\sigma_\mathrm{m\perp_1\perp_2}$ signatures over broad frequency ranges. Adjacent waves (c and d) display moderate $\sigma_\mathrm{m\perp_1\perp_2}$, while turbulent waves (e and f) maintain stochastic $\sigma_\mathrm{m\perp_1\perp_2}$ distributions. These results demonstrate the effectiveness of the wave identification criteria in distinguishing the three wave categories.

\section{Theoretical predictions for the reduced magnetic helicity of KAWs}

\begin{figure}[h!]
\centering
\includegraphics[width=7.5 in]{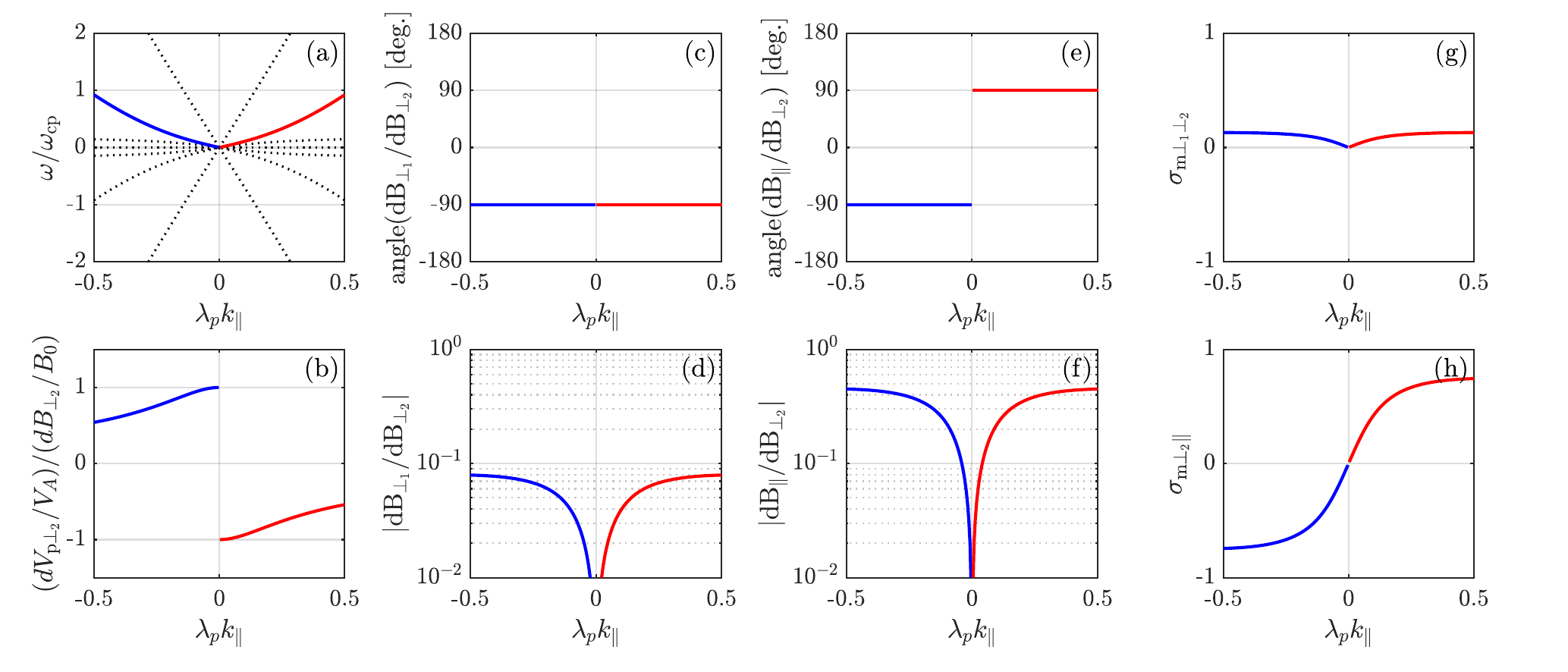}
\caption{ Properties of KAWs with wave normal angles $\theta=80^\circ$ (red) and $100^\circ$ (blue) in the plasma rest frame. (a) The wave frequency $\omega$ normalized by the proton cyclotron frequency $\omega_\mathrm{cp}$, where the dotted curves show eigenmode waves beyond the two KAWs of interest. (b) The ratio of $dV_\mathrm{p\perp_2}/V_A$ to $dB_\mathrm{\perp_2}/B_0$. (c) and (d): The argument and absolute value of $dB_{\perp_1}/dB_{\perp_2}$. (e) and (f): The argument and absolute value of $dB_{\parallel}/dB_{\perp_2}$. (g) and (h): $\sigma_\mathrm{m\perp_1\perp_2}$ and $\sigma_\mathrm{m\perp_2 \parallel}$.
}
\label{fig:KAW_rest}
\end{figure}

\begin{figure}[h!]
\centering
\includegraphics[width=7.5 in]{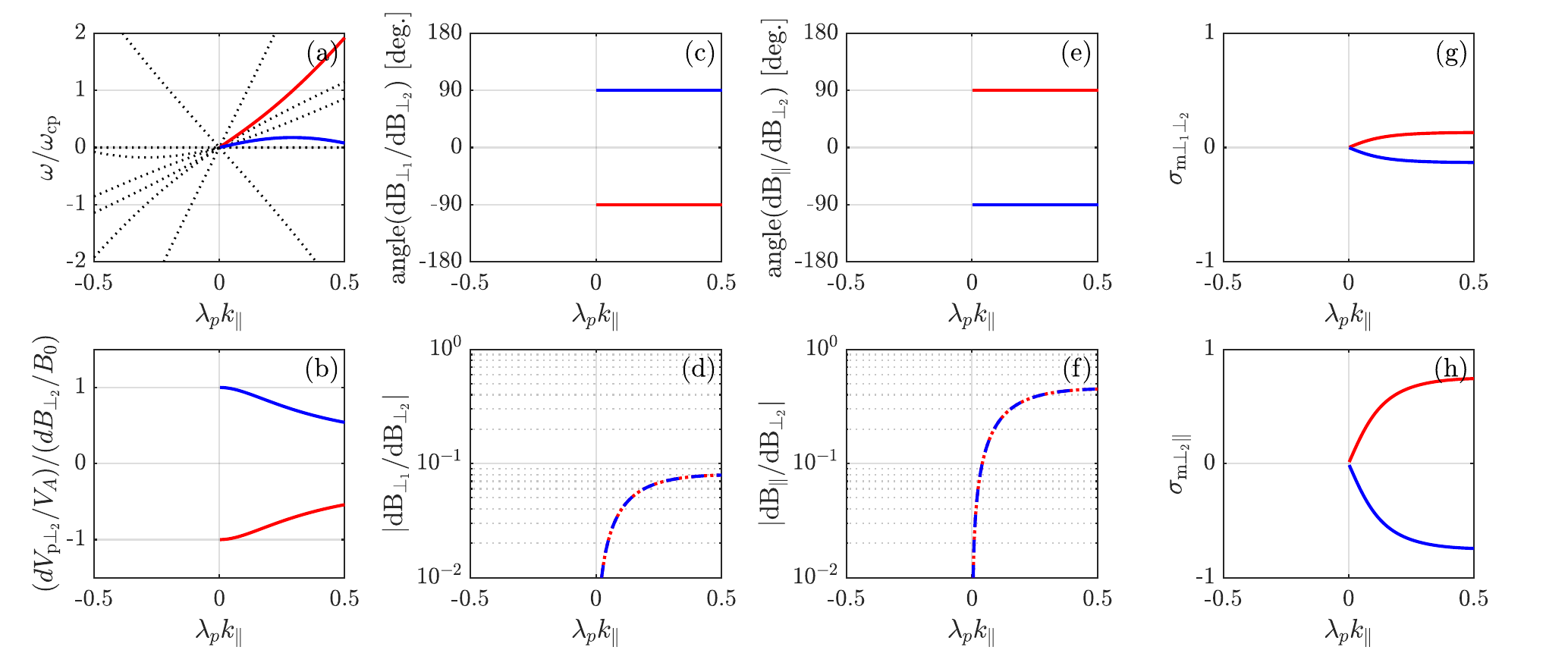}
\caption{ Properties of KAWs in the spacecraft frame, where the solar wind streams at a speed of $|{\boldsymbol V_0}|=2V_A$ along the magnetic field. The red and blue curves represent the waves with $\theta=80^\circ$ and $100^\circ$ in the plasma rest frame. The descriptions of the figure are the same as those shown in Figure \ref {fig:KAW_rest}.
}
\label{fig:KAW_outward}
\end{figure}

\begin{figure}[h!]
\centering
\includegraphics[width=7.5 in]{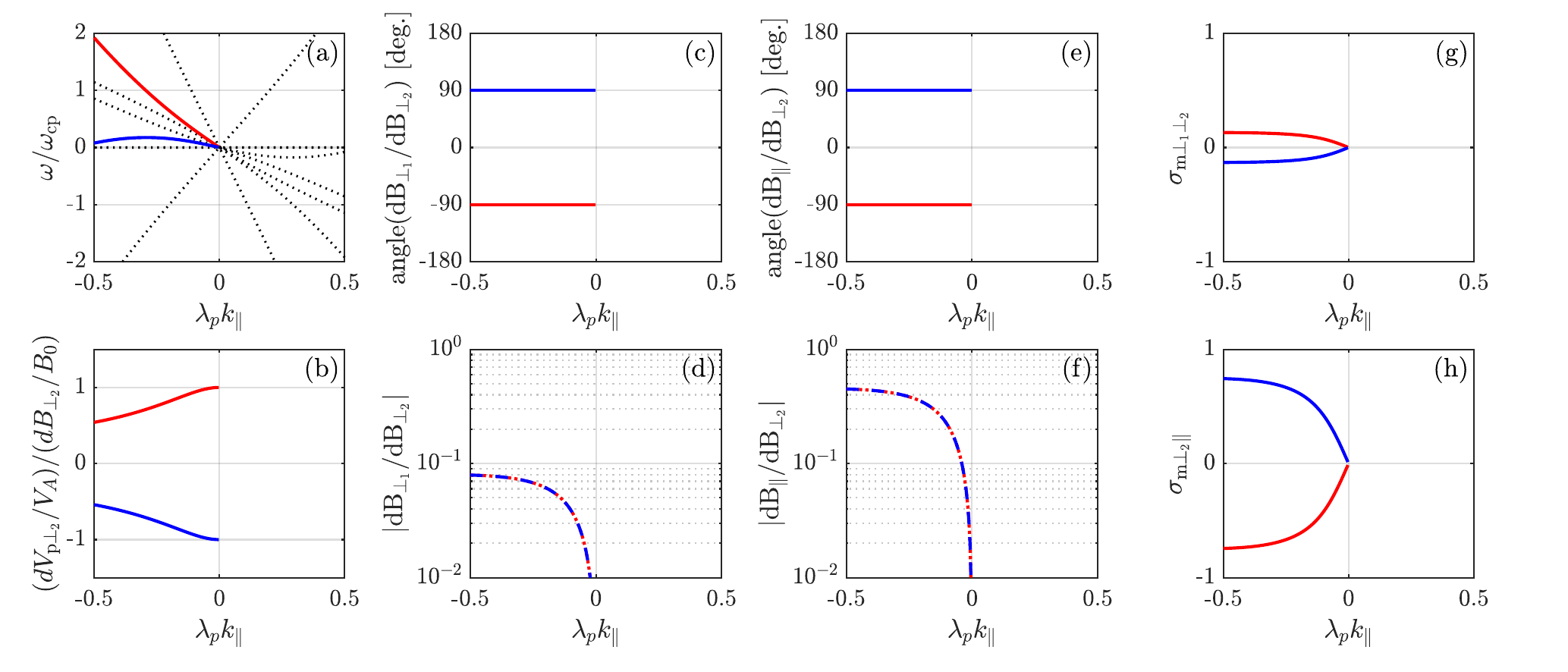}
\caption{ Properties of KAWs in the spacecraft frame, where the solar wind streams at a speed of $|{\boldsymbol V_0}|=2V_A$ against the magnetic field. The red and blue curves represent the waves with $\theta=80^\circ$ and $100^\circ$ in the plasma rest frame. The descriptions of the figure are the same as those shown in Figure \ref {fig:KAW_rest}.
}
\label{fig:KAW_inward}
\end{figure}

We use the two-fluid model to examine the magnetic helicity $\sigma_\mathrm{m\perp_2\parallel}$ of KAWs. The plasma and magnetic field parameters used are as follows: the number density $n_p=n_e=385$ cm$^{-3}$, the proton temperature $T_p=15$ eV, the electron temperature $T_e=29$ eV, and the magnetic field strength $B_0=86$ nT. The wave properties in the plasma rest frame and in the two typical spacecraft frames are illustrated in Figures \ref{fig:KAW_rest}--\ref{fig:KAW_inward}. We note that the wave properties predicted by the plasma kinetic model under these conditions are qualitatively consistent with those from the two-fluid model. A key advantage of using the two-fluid model is that it allows us to easily identify waves with the conventional definition of wave frequency $\omega$ being positive.

In the plasma rest frame, the magnetic helicity $\sigma_\mathrm{m\perp_2\parallel}$ is positive for KAWs propagating along ${\boldsymbol B}_0$ and negative for those propagating against ${\boldsymbol B}_0$, as depicted in Figure \ref{fig:KAW_rest}. The positive $\sigma_\mathrm{m\perp_2\parallel}$ has been previously discussed in the literature \citep{Gary1986JPlPh,Zhao2014ApJ}. The differing signs of $\sigma_\mathrm{m\perp_2\parallel}$ for KAWs traveling in opposite directions can be understood through the relation 
$\frac{d B_{\perp_1}}{d B_\parallel} = -\frac{k_\parallel}{k_{\perp_1}}$, 
derived from the condition $\nabla \cdot {\boldsymbol {dB}} = 0$, where ${\boldsymbol {dB}} = d B_{\perp_1} \hat{\boldsymbol e}_{\perp_1} + d B_{\perp_2} \hat{\boldsymbol e}_{\perp_2} + d B_\parallel \hat{\boldsymbol e}_{\parallel}$. Here, $d B_{\perp_1}$, $d B_{\perp_2}$, and $d B_\parallel$ represent the magnetic perturbations of KAWs in the $\hat{\boldsymbol e}_{\perp_1}$, $\hat{\boldsymbol e}_{\perp_2}$, and $\hat{\boldsymbol e}_{\parallel}$ directions, respectively. For KAWs propagating along ($k_\parallel > 0$) and against ($k_\parallel < 0$) the background magnetic field, $d B_\parallel$ exhibits different correlations with $d B_{\perp_1}$. Since $d B_{\perp_1}$ shares the same correlation with $d B_{\perp_2}$ for counter-propagating  KAWs, the correlation between $d B_{\perp_2}$ and $d B_\parallel$ varies based on the propagation direction of the KAWs.

Given that the solar wind speed is much greater than the local Alfvén speed, both outward and inward low-frequency waves are convected outward by the solar wind flow.

In the outward sector of the solar wind magnetic field, where the magnetic field points away from the Sun, outward KAWs maintain their propagation direction in the spacecraft frame. This implies that the correlations of $d B_{\perp_2}$ with $d B_\parallel$ and $d B_{\perp_1}$ remain the same. Conversely, for inward KAWs in the plasma rest frame, their propagation direction is reversed in the spacecraft frame, altering the correlation between $d B_\parallel$ and $d B_{\perp_1}$. Furthermore, this change in wave direction results in a reversed correlation between $d B_{\perp_1}$ and $d B_{\perp_2}$, as illustrated in Figure \ref{fig:KAW_outward}. Consequently, the correlation between $d B_{\perp_2}$ and $d B_\parallel$ for inward KAWs remains consistent in both the plasma rest frame and the spacecraft frame.

The changes in the correlations among the three magnetic field components of KAWs in the inward sector (where the solar wind magnetic field points toward the Sun) mirror those in the outward sector, as shown in Figure~\ref{fig:KAW_inward}. 

\begin{figure}[t]
\centering
\includegraphics[width=5.5 in]{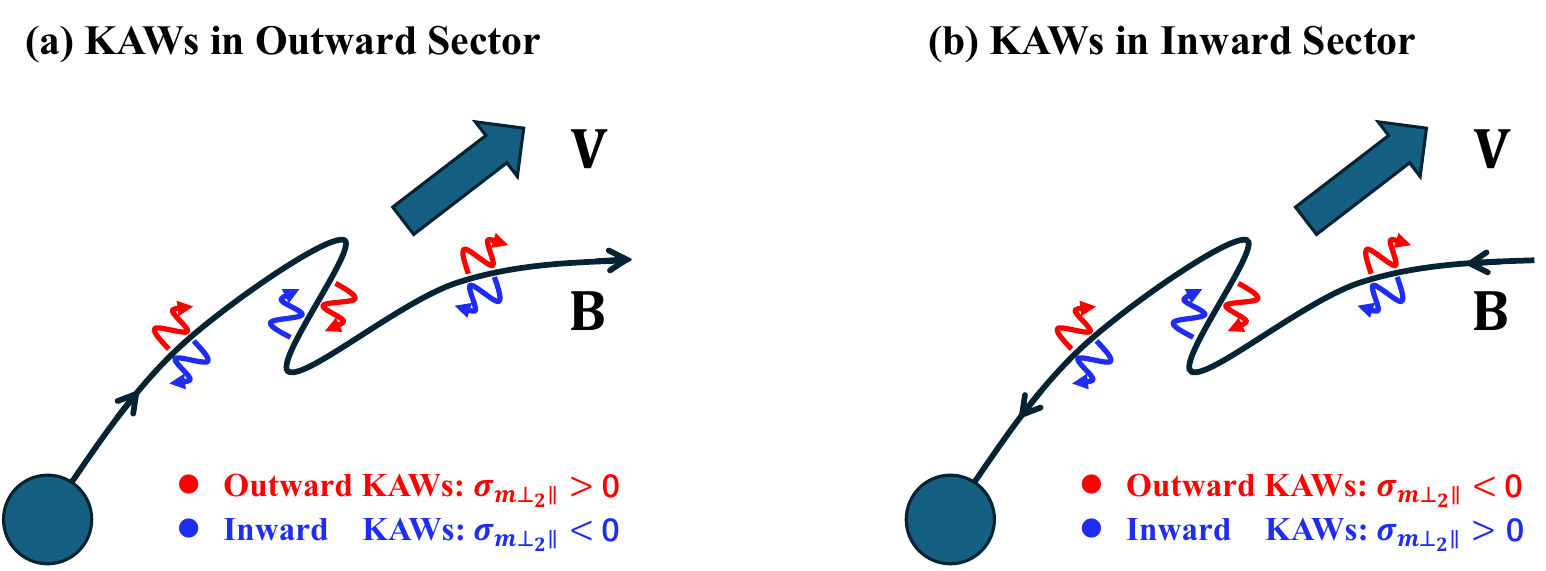}
\caption{ A schematic illustration of the sign of $\sigma_\mathrm{m\perp_2\parallel}$ for KAWs in the solar wind in the presence of switchback structures.
}
\label{fig:switchback}
\end{figure}

In conclusion, the values of $\sigma_\mathrm{m\perp_2\parallel}$ for inward and outward KAWs in the plasma rest frame are equivalent to those of the corresponding waves in the spacecraft frame. An observational confirmation of this theoretical prediction is that the sign of $\sigma_\mathrm{m\perp_2\parallel}$ remains consistent between switchback structures and the surrounding plasma environment, as shown in Figure~\ref{fig:switchback_overview}. A schematic illustration of this observation is provided in Figure~\ref{fig:switchback}.

\begin{figure}[t]
\centering
\includegraphics[width=7.0 in]{Appendix_switchback.png}
\caption{ 
(a)--(b) Distributions of (a) ${\bar N}(\sigma_{\mathrm{m}\perp_2\parallel},f)$ and (b) $R(f)$ for turbulent waves observed during magnetic field reversal intervals of switchbacks, where only reversal intervals lasting longer than 20 s are included. Data bins with durations shorter than 10 s are excluded in panel (a). (c)--(e) An example of an interval containing switchbacks (05:30–08:30 UTC on 5 November 2018), showing (c) the radial magnetic field $B_{r}$ and the magnetic helicity components (d) $\sigma_{\mathrm{m}\perp_1\perp_2}$ and (e) $\sigma_{\mathrm{m}\perp_2\parallel}$.
}
\label{fig:switchback_overview}
\end{figure}

In addition, based on Figures \ref{fig:KAW_rest}--\ref{fig:KAW_inward}, Table 1 summarizes the typical properties of KAWs in both the plasma rest frame and the spacecraft frame. To identify the propagation direction of KAWs in the plasma frame, both $\mathrm{Cor}(\delta V_{\perp2}, \delta B_{\perp2})$ (the correlation between the perpendicular velocity and magnetic field perturbations) and $\sigma_\mathrm{m\perp_2\parallel}$ serve as reliable diagnostic indicators.

\begin{table}[t]
\centering
\caption{Typical properties of KAWs in the plasma rest frame (PRF) and in the outward sector (OutSec) and inward sector (InSec) in the spacecraft frame. Here, $\mathrm{Cor}(dV_{\perp_2},dB_{\perp_2})$ denotes the correlation between $dV_{\perp_2}$ and $dB_{\perp_2}$, and $k_\parallel>0$ and $k_\parallel<0$ represent the waves propagating along and against the background magnetic field in the PRF, respectively.
}
\setlength{\tabcolsep}{0.15cm}
\begin{tabular}{ccccccl}
\hline \hline
\text{Parameters} 
& \text{OutSec:} $k_\parallel>0$ & \text{OutSec:} $k_\parallel<0$ 
& \text{InSec:} $k_\parallel>0$ & \text{InSec:}  $k_\parallel<0$ 
& \text{PRF:} $k_\parallel>0$ & \text{PRF:} $k_\parallel<0$ \\
\hline
Direction Relative to Sun
& outward & inward & inward & outward & -- & ~~~~ -- \ 
\\
Cor($dV_{\perp_2}$, $dB_{\perp_2}$) \  
&  negative  & positive  
&  negative  & positive 
&  negative  & ~~positive \
\vspace{1mm} \\ 
$\sigma_{m\perp_1\perp_2}$   \  
&  positive  & negative
&  negative  & positive
&  positive  & ~~positive \
\vspace{1mm} \\
$\sigma_{m\perp_2\parallel}$ \  
&  positive  & negative
&  positive  & negative
&  positive  & ~~negative \
\vspace{0.5mm} \\
\hline  \hline 
\end{tabular}
\label{table:expressions}
\end{table}


\end{document}